%% file: proposal.tex
\documentclass[11pt,xcdraw,table]{article}
\usepackage{amssymb}
\usepackage{graphicx}
\usepackage{changepage}
\usepackage[xcdraw]{xcolor}
\usepackage{tabularx}
\let\OLDitemize\itemize
\renewcommand\itemize{\OLDitemize\addtolength{\itemsep}{0pt}}
\usepackage{url}
\usepackage{enumitem}
\setitemize{itemsep=-4pt}
\usepackage{geometry}
\usepackage{booktabs}
\usepackage{wrapfig}
\usepackage{multirow}

\providecommand{\keywords}[1]
{
	\small	
	\textbf{\textit{Keywords---}} #1
}

\title{DF-Captcha: A Deepfake Captcha for Preventing Fake Calls \\
	\vspace{1em}
	\large{} A draft academic paper based on the provisional patent submitted January 1st 2022 under provisional Number 63/302,086
	\\
	\vspace{1em}
	\flushleft \normalsize{}
	\vspace{-2em}
}

\author{Yisroel Mirsky\\Offensive AI Research Lab\\Ben-Gurion University\\yisroel@post.bgu.ac.il\\\url{https://offensive-ai-lab.github.io/}}
\begin{document}

\clearpage\maketitle
\pagestyle{plain}

%\newpage
%\tableofcontents
%\newpage

\begin{abstract}
	Social engineering (SE) is a form of deception that aims to trick people into giving access to data, information, networks and even money. For decades SE has been a key method for attackers to gain access to an organization, virtually skipping all lines of defense. For example, numerous organizations have been infiltrated through spear phishing attacks where targeted emails were written to entice the receiver to click on a malicious link or open an infected attachment. Attackers also regularly use SE to scam innocent people by making threatening phone calls which impersonate an authority or by sending infected emails which look like they have been sent from a loved one. 
	SE attacks will likely remain a top attack vector for criminals because humans are the weakest link in cyber security. 
	
	Unfortunately, the threat will only get worse now that a new technology called deepfakes as arrived. A deepfake is believable media (e.g., videos) created by an AI. Although the technology has mostly been used to swap the faces of celebrities, it can also be used to `puppet' different personas. Recently, researchers have shown how this technology can be deployed in real-time to clone someone's voice in a phone call or reenact a face in a video call. Given that any novice user can download this technology to use it, it is no surprise that criminals have already begun to monetize it to perpetrate their SE attacks. 
	
	In this paper, we propose a lightweight application which can protect organizations and individuals from deepfake SE attacks. Through a challenge and response approach, we leverage the technical and theoretical limitations of deepfake technologies to expose the attacker. Existing solutions are too heavy as an end-point solution and can be evaded by a dynamic attacker. In contrast, our approach is lightweight and breaks the reactive arms race, putting the attacker at a disadvantage.

\end{abstract}
\keywords{Deepfakes, social engineering, phishing, cybersecurity}

\newpage

\section{Introduction}
%light intro, what has been done in the past, what the gap is. lots of cites.

A deepfake is content, generated by an artificial intelligence, that is authentic in the eyes of a human being. The word \textit{deepfake} is a combination of the words \textit{`deep learning'} and \textit{`fake'} and primarily relates to content generated by an artificial neural network, a branch of machine learning.

The most common form of deepfakes involve the generation and manipulation of human imagery. 	
This technology has creative and productive applications. For example, realistic video dubbing of foreign films,\footnote{https://variety.com/2019/biz/news/\\ai-dubbing-david-beckham-multilingual-1203309213/} education though the reanimation of historical figures \cite{Deepfake69:online}, and virtually trying on clothes while shopping.\footnote{https://www.forbes.com/sites/forbestechcouncil/2019/05/21/gans-and-deepfakes-could-revolutionize-the-fashion-industry/} There are also numerous online communities devoted to creating deepfake memes for entertainment,\footnote{https://www.reddit.com/r/SFWdeepfakes/} such as music videos portraying the face of actor Nicolas Cage.

\begin{wrapfigure}{r}{0cm}
	\centering
	\includegraphics[width=0.5\textwidth]{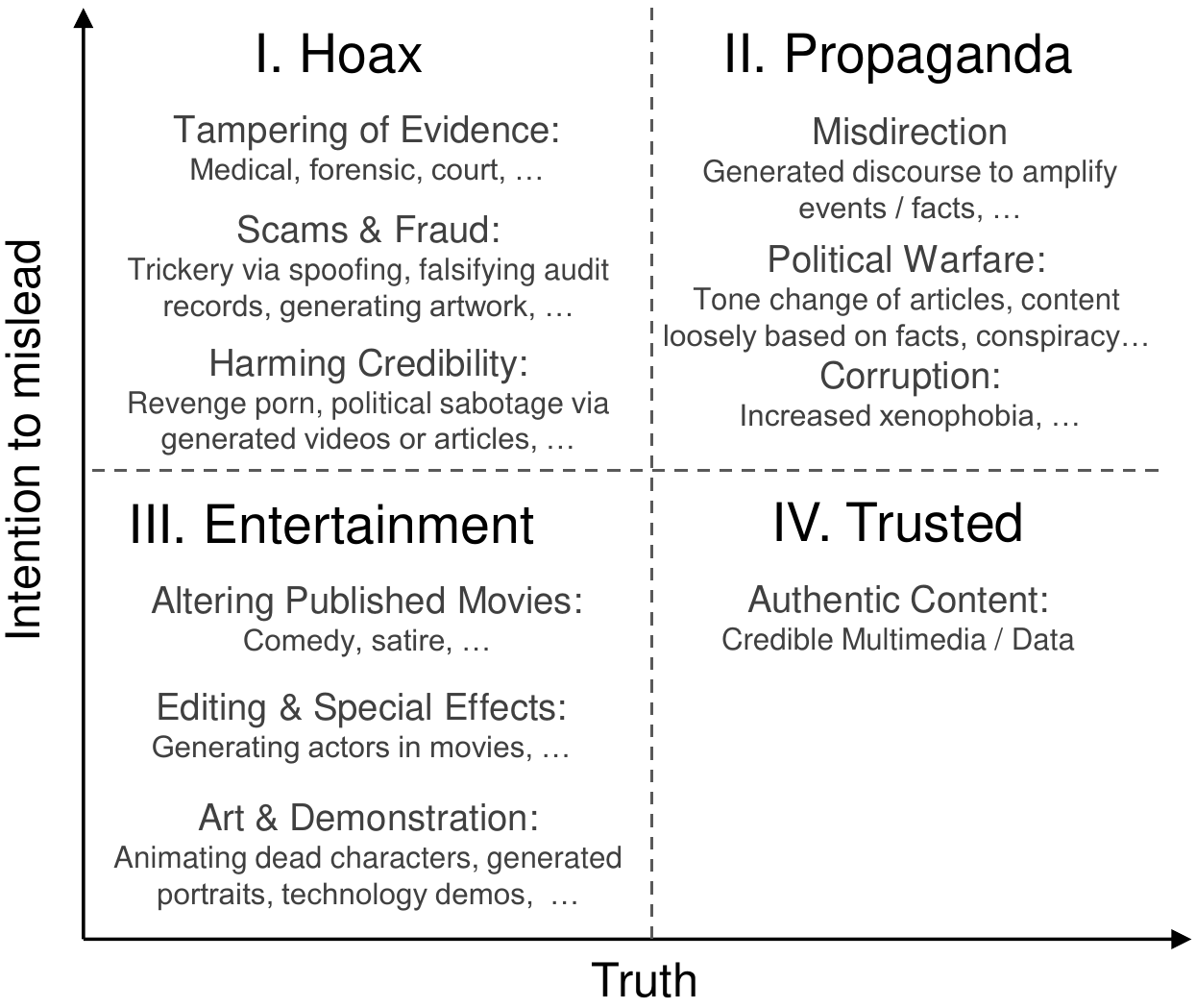}
	\caption{A deepfake information trust chart.}\label{fig:infochart}
\end{wrapfigure}
However, despite the positive applications of deepfakes, the technology is infamous for its unethical and malicious aspects. 
At the end of 2017, a Reddit user by the name of `deepfakes' was using deep learning to swap faces of celebrities into pornographic videos, and was posting them online\footnote{https://www.vice.com/en\_us/article/gydydm/gal-gadot-fake-ai-porn}.
The discovery caused a media frenzy and a large number of new deepfake videos began to emerge thereafter. In 2018, BuzzFeed released a deepfake video of former president Barak Obama giving a talk on the subject. The video was made using the Reddit user's software (FakeApp), and raised concerns over identity theft, impersonation, and the spread of misinformation on social media. 

Overall, there are a number of different uses for deepfakes. However, many of them are criminal and unethical. Fig. presents an information trust chart which summarizes how deepfakes can be used to mislead people. In this paper, we are concerned with hoaxes (the upper left quadrant).%maybe mention how damage is also in that people dont' trust anything paper. The threat of deepfakes has 

%Today, deepfakes have become hyper realistic (see Fig. \ref{fig:realism}) and easily accessible through free open source repositories that can be run with little experience. For example, there are a variety of free user-friendly face swapping tools which can be used to reenact a target []. However, what more alarming are the realtime.. and voice cloning using just 3 seconds of audio [].}

\subsection{The Threat} Criminal activities usually motivated with monetary gain. Therefore, in the coming years we can expect to see deepfakes being weaponized for monetization. This even more likely considering that the technology has already been proven as an effective attack tool for humiliation, misinformation, and defamation. Moreover, deepfake technologies are becoming more practical \cite{deepfake32:online}, efficient \cite{jia2018transfer}, and easily accessible.\footnote{https://github.com/aerophile/awesome-deepfakes}

The most likely scenario is that deepfakes will be used in social engineering (SE) attacks. An SE attack is where an attacker uses psychological manipulation to trick users into making security mistakes or giving away sensitive information. For example, enticing a user to open an email attachment or transfer money in an phone scam. 

SE is so effective that 98\% of all cyber attacks rely on it \cite{2020Cybe88:online}. This is because humans are the weakest link in cyber security. We are inherently gullible when enticed, threatened, or lured into a false pretext in a social encounter. Cyber criminals know this and exploit these vulnerabilities to achieve their goals. As a result SE attacks affect everybody: the public sector, private sector, and the individual at home.

A major concern is, what happens when criminals start using deepfakes to exploit our trust? Imagine you receive a call from your mother who is in trouble and urgently needs some money transferred. The caller sounds exactly like her, but the situation seems a bit out of place. Under stress and frustration, she hands the phone over to your father who confirms the situation. Without hesitation, many would transfer the money --even though they talking to a stranger.

\begin{figure}[t]
	\flushright
	\includegraphics[width=0.8\textwidth]{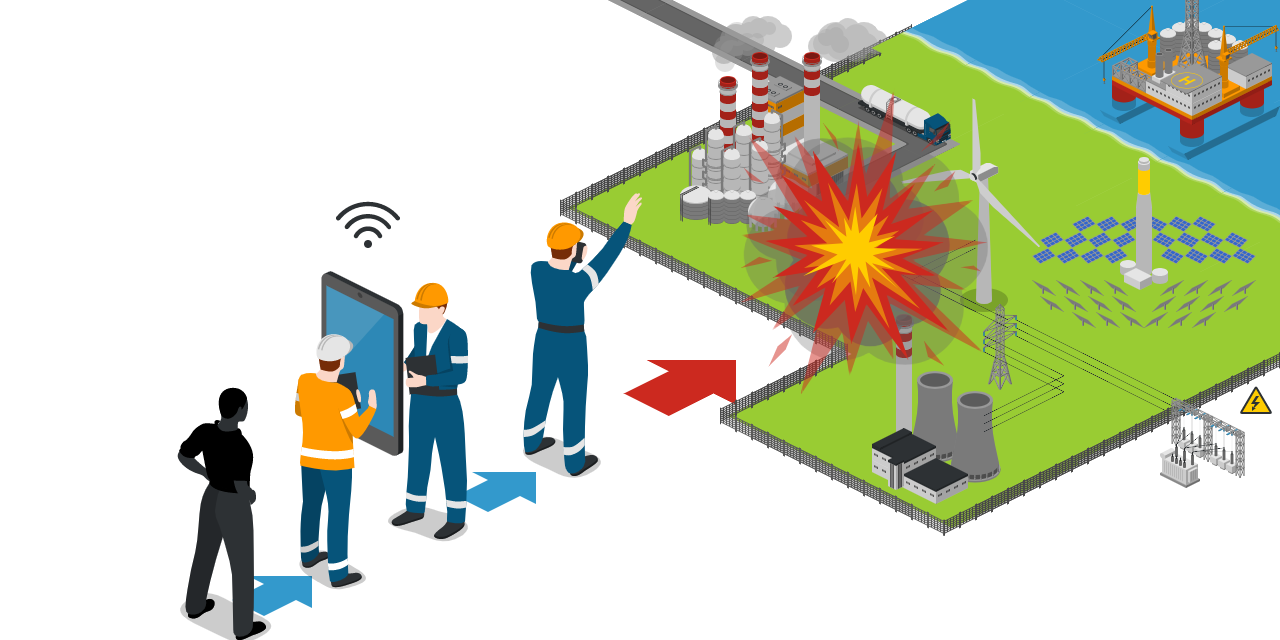}
	\caption{An illustrative example of how a deepfake social engineering attack can impact critical infrastructure.}\label{fig:critical}
\end{figure}

Now consider state-actors with considerable amounts of time and resources. They could target workers at power plants and other critical infrastructure by posing as their superiors. Over a phone call they could convince the worker to change a configuration or setting which would lead to a cyber breach or a catastrophic failure (Fig. \ref{fig:critical}). The attackers could even target soldiers or officers on duty leading to the compromise of our national security. 

\begin{figure}[bh]
	\centering
	\includegraphics[width=0.32\textwidth]{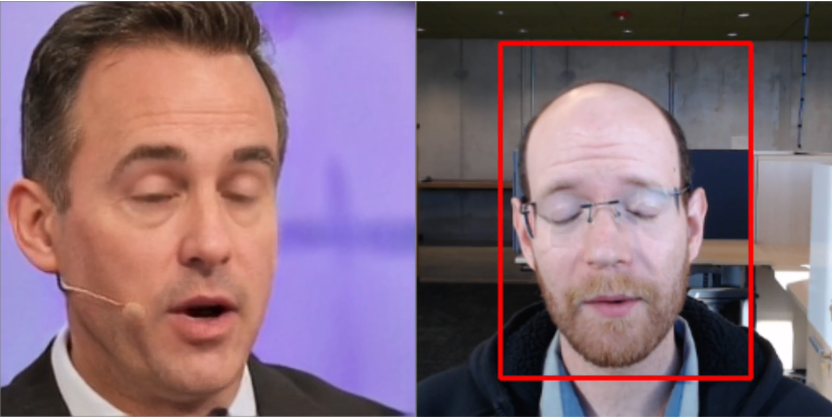}	
	\includegraphics[width=0.32\textwidth]{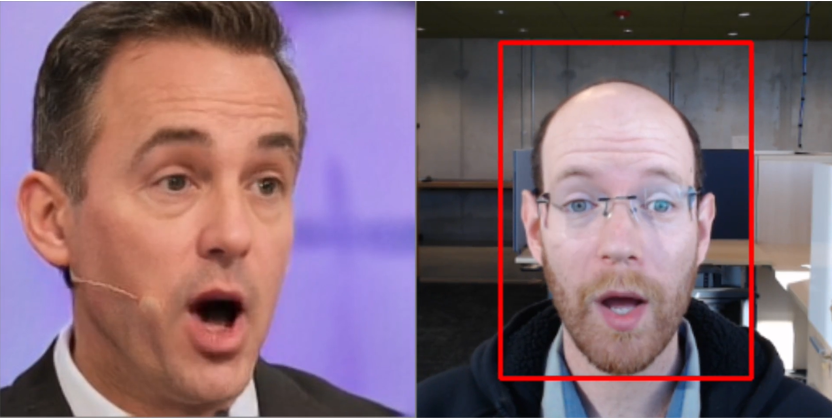}	
	\includegraphics[width=0.32\textwidth]{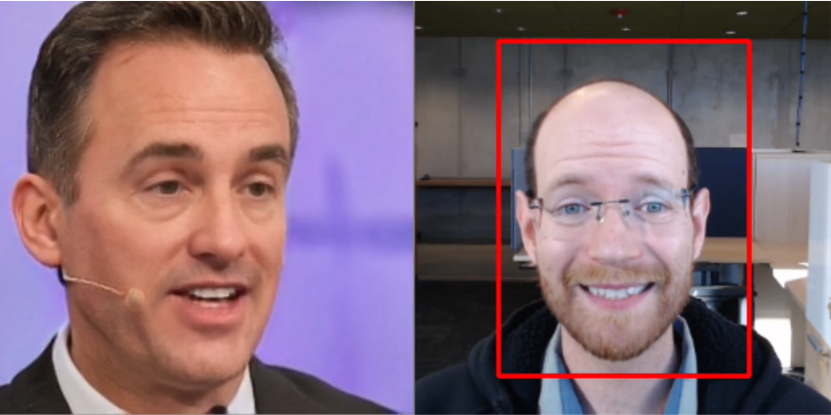}
	\caption{Still frames taken from a real-time deepfake reenactment video of a CEO. The right of each image is the driver (attacker) and the left is the resulting frame. The deepfake was generated using \cite{NIPS2019_8935} with a \textit{single} image of the CEO.}\label{fig:ceo}
\end{figure}

These kinds of SE attacks would require the rendering of deepfakes in real-time. Although most methods cannot be rendered in real-time, researchers have recently shown that this is indeed possible. For example, in  \cite{jia2018transfer} the authors show how one's voice can be cloned with only five seconds of audio.\footnote{For this paper, we have prepared an audio demo of this attack where 3 seconds of voice is used to clone a CEO's voice and use it to make a call: https://tinyurl.com/nzsszh54} Furthermore, the authors in \cite{NIPS2019_8935} propose a first order motion model which can be used to reenact (puppet) any still image in real-time (Fig. \ref{fig:ceo}).

Deepfake SE attacks are tangible threats. In 2019, a company was scammed out of \$250K when an associate was tricked over the phone to transfer money. The call was later confirmed to be deepfake of the CEO's voice \cite{AVoiceDe76:online}. After that, a widow was scammed out of nearly 300k\$ when conned in a false romance by a real-time deepfake video \cite{RomanceS61:online}. In 2020, cybercriminals stole \$35 million in a bank heist where a bank manager received a deepfake phone call in the voice of a company director \cite{Fraudste51:online}. In 2021 European MPs participated in Zoom meetings with a real-time deepfake of a Russian opposition leader \cite{European99:online}. Even more recently, in 2022 the FBI warned the public of cybercriminals which are using real-time deepfakes to perform interviews and gain access to organization in elaborate social engineering attacks \cite{FBIScamm95:online}. At the rate which these attacks are progressing, it is only a matter of time before they become common place.

\subsection{The Proposed Solution}
In this paper, we propose the DF-Captcha: a challenge-response turing test against real-time deepfake technologies. The concept is that when a victim receives a suspicious audio or video call, the victim can send the caller a challenge which can be automatically verified. What makes this technique powerful is that the challenge is easy for a human to perform but extremely hard for a deepfake model to generate. For example, to expose facial reenactment, the challenge might be to have the caller move his/her head to an oblique angle, press on the nose, or simply turn around. 

The nature of these challenges is that they target limitations in deep learning technology. As a result, models which attempt them either do not perform the action or generate large artifacts which can easily be detected. 

The limitations stem from two main challenges for the attacker:
\begin{enumerate}
	\item It is hard for the attacker to procure a dataset which captures many different non-related activities, situations, and phyisical simulations all while perfecting the related task (e.g., facial reenactment).
	\item Current deep learning technology does not multitask well (it cannot generate hyper realistic content all while managing temporal physics, material simulations, anatomy, etc.) Although it is foreseeable that the limitations may diminish as deep learning matures, the victim can easily add more challenges requiring the attacker to train every model to be an expert at every single one. This puts the defender one step ahead of the attacker.
	
\end{enumerate}

\noindent DF-Captcha has several advantages over existing defenses (described in detail in section \ref{sec:relwork}): 
\begin{description}
	\item[Robust.] It does not rely on a single artifact that can refined (past works). Moreover, it	‘forces’ the attacker to exhibit large artifacts, and the artifacts appear regardless of the deep fake's quality. 
	\item[Resilient.] This defense is hard for attacker to evade. The attacker cannot cover all concepts (tasks) with high fidelity –concepts, loss functions, etc. It is also easy for defender to introduce new challenges.
	\item[Efficient.] Only the relevant content needs to be analyzed (not every frame in the video/audio)
	\item[Flexible.] The responses can be automatically or manually verified.
\end{description}

\section{Overview of Proposed Method}

\noindent\textbf{Mitigating Real-time Deepfakes.} To keep ahead of the game, we must be proactive and consider the adversary's next step. If we just consider the weaknesses of the current attacks then the adversary can evade detection and will always have the upper hand. Instead, we suggest that an effective defense targets the adversary's limitations. With this strategy in mind, we propose a simple yet effective defense where suspicious callers must complete a `deepfake Turing test' to be verified as a real persona. The test builds on the first research objective since the challenges posed to the deepfake are explicitly designed to stress the deepfake's limitations. As a result large artifacts will be created which can be easily detected by the victim or automatically via light anomaly detection models. The core research objectives are to design a extensible set of challenges which are both non-intrusive to a legitimate caller and highly effective in exposing the real-time deepfake.

The key difference between our approach and past works is that none of the current existing works utilize or exploit the limitations of deepfakes their advantage. Doing so not only strengthens the position of the defender in detecting the attack, but also provides some guarantees on doing so as well. For example, today there does not exist any deepfake technology which can render the head at oblique poses or handle voice conversion with a wide range of accents or inflection not found in the training set of the target. Although this seems trivial, and we know that deepfakes will continue to improve, we also know that there will always be limitations which we can exploit. By forcing the attacker to reveal these limitations, strong artifacts will be presented which we will be able to easily detect with light weight models (e..g, light networks, classical machine learning such as SVMs, or statistics). As such, part of this work will be to (1) enumerate the limitations of deepfakes today and how that can be compiled into non-intrusive challenges and (2) identify the limitations of deepfakes of the future based on concrete theoretical bounds of GANs because of their dependence on their training sets (i.e., they cannot generate content or behaviors not found in their data). Furthermore, our proposed approach will alleviate the concern of a victim ignoring the warning of a detector (e.g., as a false positive). This is because we will force the attacker to reveal defects which are obvious to the victim.

The significance of the research is that it will prepare us to handle the threat of deepfakes before they become mainstream. Defense researchers will also benefit from our work by guiding their efforts according to our findings. Moreover, society will benefit from a free, extensible, open-source application which can be used to prevent these attacks on the public sector, private sector, and individuals of all ages. To the best of our knowledge, the proposed tool is the first method designed to detect real-time deepfakes, and the concept of using a deepfake Turing test is also novel.

This method can prepare society for a new and advanced social engineering attack which is emerging. Although attacks and technologies evolve overtime, this method is modular enabling the defender consider new limitations by adding new tests. Therefore the system has longevity since it can be easily extendable.

\section{Description of the Method}\label{sec:desc_re}
The proposed defense is similar to a Turing test. A Turing test is a test of a machine's ability to exhibit intelligent behavior which is indistinguishable from that of a human. These test have been used the past for cyber security. For example, to prevent myriads of bots from attacking Internet services, CAPTCHAs are used where the user (human or bot) must solve a visual in order to prove that the user is human. 

\begin{figure}[t]
	\flushright
	\includegraphics[width=0.8\textwidth]{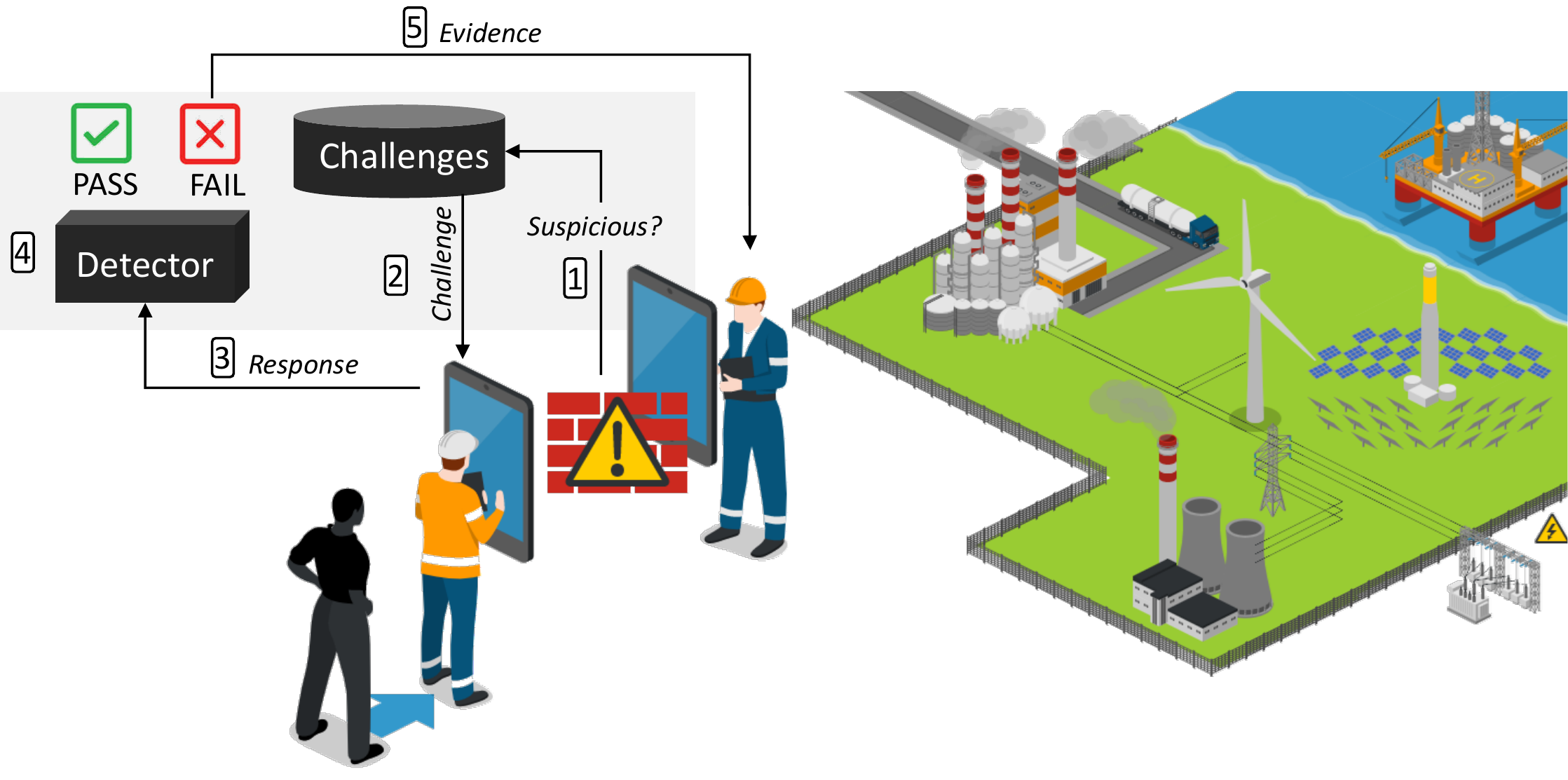}
	\caption{The process in which the proposed defense detects and protects victims from real-time deepfake SE attacks.}\label{fig:defense}
\end{figure}

In this research we propose a `deepfake Turing test' where a user is challenged to produce audio or video content, and the requested content (i.e., the challenge) is a known limitation of generative AI. Following this approach, a call screening application can be designed to help users identify deepfake calls. This defense works as follows (illustrated in Fig. \ref{fig:defense}):
\begin{enumerate}
	\item The defense is triggered if a video or audio call is suspicious. This is determined by one or more indicators configured by the user: (1) the caller is new and hasn't been verified yet, (2) the caller has indicative network history such as malicious past behaviors, (3) the caller is masking his/her true identity (e.g., the call originates from the Tor network), (4) the liveness \cite{kavita2020contemporary} of the caller is very low or other obvious artifacts, (5) the user (victim) has requested a test mid-call due to suspicious behaviors or to ensure authenticity before carrying out a caller's requests.
	\item A challenge is sent to the caller using a prerecorded message. The challenge is automatically selected based on the type of call (audio or video), the call quality, the level of suspicion, and the caller's perceived status  (e.g., indoors, outdoors, sitting, etc.)
	\item The caller performs the requested challenge and the content is captured.
	\item The detector (1) identifies and extracts the response content (i.e., the few frames or a second of audio containing the performance of the challenge) and (2) applies a light weight anomaly detector to the response content.
	\item The user is notified of the result (pass or fail) and is given (1) a degree of certainty measured by the model's confidence, and (2) the evidence in the form of the the capture challenge content. The user can then decide whether to proceed with the call,  examine the evidence closer him/herself, or request another challenge. 
\end{enumerate}

\noindent\textbf{The Challenges.} We have identified three categories of challenges which stress the limits of a model's training set and technology: physical, out-of-distribution, and audio. The following is an initial set example challenges which we have identified. These challenges produce large artifacts in real-time deepfakes and are easy for a human to perform. 

A technology challenge (or simulation challenge) aims to push the limits of a deepfake's capability. They involve activities where the action cannot be done with existing deepfake technology or cannot be done well by a model which has been optimized for reenactment (e.g., facial puppetry). Examples include: tongue motions, poke cheek or nose, fold ear, vibrate lips, crease shirt, stroke hair, show an object, remove glasses, wave hand quickly (motion blur), drop/bounce object, hold objects between fingers, and interact with a select object in the background.

\noindent Examples of technology challenges:
\begin{enumerate}
	\item Drop object
	\item Bounce Object
	\item Fold shirt
	\item Stroke hair
	\item Interact with background scenery
	\item Spill water
\end{enumerate}

An out-of-distribution (OOD) challenge aims to exploit the limited training set of the attacker's model. If the attacker has limited data on the victim (i.e., less than 60 hours of video from all angles and expressions) then the model will struggle to extrapolate the spatial content. This also holds true when using zero or few-shot learning \cite{lee2019metapix,zakharov2019few,wang2019fewshotvid2vid,shaoanlu29:online} where a generalized model trained on many other people's data is fine tuned on the target. For example, stand up, move close to camera, move head to oblique angles, leave screen, perform occlusions (e.g., hold hand in front of face), open mouth, and perform hand expressions.

\noindent Examples of OOD challenges:
\begin{enumerate}
	\item Pick up requested object
	\item Hand expressions
	\item Tongue motion
	\item Fold ear
	\item Face occlusions
	\item Remove glasses
\end{enumerate}

An audio challenge is designed to perform a technological or OOD test on the voice of the caller. For example, sing a few notes, repeat phrase with the given rhythm, use a different accent, talk with a given tone or speed, speak very close to microphone, whisper, and clear throat.

\noindent Examples of technology challenges:
\begin{enumerate}
	\item Mimic phrase 
	\item hum tune
	\item Sing part of song
	\item Repeat accent
	\item Change tone or speed
	\item Clear throat
	\item Whistle
\end{enumerate}

The tool can be extended to include more challenges as time goes on.

\noindent\textbf{The Detector.}
The objective of the detector is to locate and evaluate the response of the caller. To make the detector efficient and practical, we perform this process in two steps. First the response is extracted from the captured content in the form of a short audio clip or sequence of frames $R$. The coarse location of the response can be located by searching for the anticipated activity with a light weight machine learning tool. For example, to find when the head is turned to an oblique angle we can predict the caller's pose using open source software like \cite{osokin2018real}. Next, $R$ is passed through an anomaly detector trained on clean content of other people performing the same challenge. 
%how get thsi dataset?
Finally, the model predicts whether $R$ is legitimate or not and then passes $R$ with its confidence score to the user.

We are not concerned that an attacker may use adversarial machine learning \cite{chakraborty2018adversarial} to trick our models. This is because (1) the media channels are lossy due to compression which is an effective mitigation against these techniques, and (2) the work the attacker could do cause the detector to retrieve the wrong content as $R$ which will result in an anomaly anyways. 

For the anomaly detection model we will experiment with a variety of algorithms and architectures to find the best trade-off between performance and efficiency (speed and resources). Initially, we will utilize our experience in lightweight neural networks for anomaly detection to perform the task \cite{mirsky2018kitsune,meidan2018n}. We will also try using once-class SVMs and statistical models as a baseline. We expect that these light models will perform very well given that we know the exact location of the response and that the signals from artifacts will be very strong. Lastly, the set of challenges are extensible making it easy to add new ones as novel deepfake technologies are released and new limitations are exposed.

\section{Preliminary Results}
\begin{figure}[t]
	\centering
	\includegraphics[width=\textwidth]{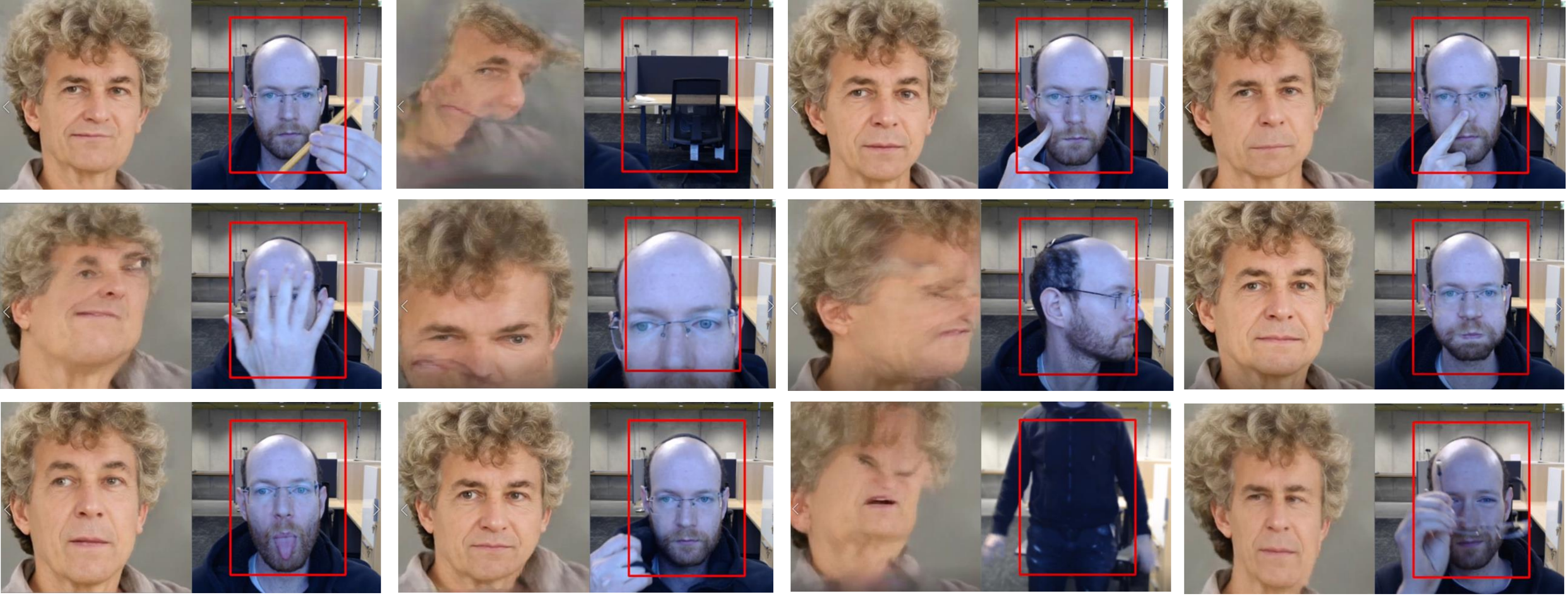}
	\caption{Preliminary results demonstrating the weaknesses of real-time deepfakes to various challenges.}\label{fig:prelim}
	\vspace{-1em}
\end{figure}

We implemented the real-time deepfake presented in \cite{NIPS2019_8935} and reenacted a target photo. We achieved a realistic live video deepfake at 35 fps. Next, we performed a sample of the challenges listed in section \ref{sec:desc_re}. Fig. \ref{fig:prelim} presents some screenshots of these challenges. The preliminary results show that our intuition is correct: deepfake attacks are limited and can be easily detected when forced to perform a deepfake Turing test.  

\section{Strengths of the Method}
%arms race, can we break it?

%rq and hypothsis
%Closing the Researhc gap

In this research project, we aim to address the following research questions:
\begin{itemize}
	\item This defense will effectively mitigate the threat since it is much easier for the defender to develop and add new challenges then it is for the attacker to develop better deepfake generative technology. 
	\item Since a realistic deepfake must focus its training effort of facial quality, there exists a significant range of poses, expressions, actions, and simulations which cannot all be captured by the generative model. Therefore, there is a high likelihood that a sufficiently sized subset of these challenges will be easy for a human to perform.
	\item Detection is practical because it is possible to limit the detection to moments which most likely contain the response, thus reducing the number of false positives. Moreover, by using anomaly detection, we can use a single model for all existing and new challenges, making the detector practical. Finally, by leveraging the weaknesses of the attacker's deepfake model we can increase the probability of detecting the artifacts making the approach accurate.
	\item Attackers are vulnerable because generative AI cannot create realistic content for expressions that do not appear in the training set (e.g., generate the back of the head, or the victim's singing voice). This is because there are a wide variety of expressions and interactions and it is extremely challenging for an attacker to collect all of these aspect with the victim. Moreover, generalizing with media from other people will still require the model to extrapolate the missing information leading to blurry or inconsistent content.
	\item Even if the attacker has lots of media on the target (and can make a hyper realistic deepfake of him/her) the defense will work well. This is because some of the challenges are not data dependent, but rather technology dependent (e.g., they stress the limits of occlusions, temporal coherence, physics, synthesis of arbitrary objects, etc.) 
	
\end{itemize}

\section{Related Work - Current Countermeasures}\label{sec:relwork}
Defenses against deepfakes have been a subject research in the scientific community for some time. A summary and systematization of the deepfake detection methods can be found in Table \ref{tab:detection}. The primary goal of these works have been to identify deepfakes by either finding artifacts or by using generic classifiers.

%artifact specific
\subsection{Artifact-based Detection}
Deepfakes often generate artifacts which may be subtle to humans, but can be easily detected using machine learning. In 2014, researchers had this hypothesis and monitored physiological signals, such as heart rate, to detect computer generated faces \cite{conotter2014physiologically}. Regarding deepfakes, \cite{li2018ictu} monitored irregular eye blinking patterns and \cite{ciftci2019fakecatcher} monitored blood volume patterns (pulse) under the skin. 

Inconsistencies are also a revealing factor. In \cite{korshunov2018speaker} and \cite{korshunov2019tampered}, the authors noticed that video dubbing attacks can be detected my correlating the speech to landmarks around the mouth. In \cite{yang2019exposing} it was shown that similar artifacts appear when predicting the facial landmarks. With large amounts of data on the target, mannerisms and other behaviors can be monitored for anomalies. For example, in \cite{agarwal2019protecting}	the authors protect world leaders from a wide variety of deepfake attacks by modeling their recorded stock footage.

Some artifacts appear where the generated content was blended back into the frame. The authors of \cite{agarwal2017swapped,zhang2017automated,akhtarcomparative,mo2018fake,durall2019unmasking} use edge detectors, quality measures, and frequency analysis to detect artifacts in the pasted content and borders. In, \cite{li2019face} the authors detect deepfakes by decomposing them into to their sources while identifying the content's boundary. Some works identify and visualize the tampered regions by either predicting masks learned from a ground truth, or by mapping the neural activations to the raw image   \cite{nguyen2019multi,du2019towards,stehouwer2019detection,li2019zooming}.

The content of a fake face can be anomalous in context to the rest of the frame. For example, residuals from face warping processes \cite{li2019exposing,danmohah54:online,li2019celeb}, lighting \cite{straub2019using}, and varying fidelity \cite{korshunov2018deepfakes} indicate the presence of generated content.
In \cite{yu2019attributing} and \cite{marra2019gans}, the authors found that GANs leave unique fingerprints and show how it is possible to classify the generator given the content, even in the presence of compression and noise. In \cite{koopman2018detection} the authors analyze a camera's unique sensor noise (PRNU) to detect pasted content.

Realistic temporal coherence is challenging to generate, and some authors capitalize on the resulting artifacts to detect the fake content. For example, \cite{guera2018deepfake} uses an RNN to detect artifacts such as flickers and jitter, and \cite{sabir2019recurrent} uses an LSTM on the face region only. In \cite{amerini2019deepfake} the optical flow between frames is analyzed, and in \cite{chan2019everybody} a classifier is trained on the two frames directly.\looseness=-1

Most works on voice clone detection fall under this category as well. Unfortunately, very little work has been done to detect deepfake voices. Previous works have focused on voice synthesized using non-deep learning approaches \cite{albadawy2019detecting}. As a result, these methods have trouble identifying the same artifact found made by deep learning models. This is because deep learning models are able to generate fluid audio and because the artifacts lie in other feature sets (e.g., behavior). 

The first detection model for deepfake audio was published in 2019 \cite{albadawy2019detecting}. There the authors proposed a bispectral analysis method for detecting AI-synthesized fake voices.
They observed that specific and unusual spectral correlation exhibited in the fake voices synthesized with DNNs. They explore these bispectral artifacts using higher-order polyspectral
features for discriminating fake voices. Other works use deep learning models, such as recurrent neural networks, in a generic way to classify the audio samples \cite{chintha2020recurrent,chen2020generalization}. Finally, in \cite{wang2020deepsonar} the authors propose DeepSonar which observes the neural activation of a voice classification network. If the activation pattern appears anomalous, then the voice sample is considered fake. 

\subsection{Undirected Approaches} Instead of focusing on a specific artifact, some authors train deep neural networks as generic classifiers, and let the network decide which features to analyze \cite{afchar2018mesonet,do2018forensics,tariq2018detecting,ding2019swapped,fernando2019exploiting}. In \cite{marra2018detection,rossler2019faceforensics++,nguyen2019capsule}, it was shown that deep neural networks tend to perform better than traditional image forensic tools on compressed imagery. Alternatively, to overcome noise and other distortions, the authors of \cite{wang2019fakespotter} measure the neural activation (coverage) of a face recognition network to obtain a stronger signal from than just using the raw pixels. In \cite{bao2018towards} the authors detect deepfakes by measuring an input's embedding distance to real samples using an ED's latent space. To improve performance of their models, some add noise to their training data \cite{xuan2019generalization} and others try to develop models which generalize better to unseen attacks/generators through disentanglement \cite{cozzolino2018forensictransfer} and semi-supervised learning \cite{tu2019deep}.

\subsection{Challenge Response Approach}
In \cite{uzun2018rtcaptcha} the authors propose rtCaptcha. In this paper the authors detect fake video calls by assuming that attackers cannot generate content fast enough. They have the user speak a written captcha and rely on the assumption that the attacker will not be able decipher it and then pass it through his voice generator in time. This approach works against attackers who are using still images of the victim, prerecorded audio/video recordings of the victims, and older deepfake technologies which required offline production. However deepfake technology has advanced to the level where high quality attacks can be performed in real-time audio and video are processed per frame with negligible reaction delay --bypassing rtCaptcha completely. Instead, we aim to identify real-time deepfake attacks through a captcha which exploits the weaknesses a real-time deep fake technology: making it hard for the attacker and easy for the defender to keep up.

\input{detection_table.tex}

\subsection{The Scientific Gap}
The issue with current approaches is that they are not suitable for detecting real-time deepfake attacks. This is because they (1) cannot be run in real-time, (2) cannot be run on a smartphone (where we expect an SE attack to be received), (3) assume a clean lossless channel \cite{chen2020generalization,chintha2020recurrent,wang2020deepsonar} which is not the case for phone calls or VoIP call (whatsapp etc) where compression rates can vary based on connectivity, and (4) do not handle the human side of the issue, where a victim may ignore the warnings due to the urgency of the attacker under the false pretex.
However, the most fundamental issue is that all of these methods perpetuate an endless arms race. This is because for every artifact or detector, the adversary can tune his/her model to evade detection, putting the defender at a disadvantage.

\vspace{.5em}\noindent\textbf{Evading Artifact-based Detectors.} 
To evade an artifact-based detector, the adversary only needs to mitigate a single flaw to evade detection. For example, the attacker's model can generate the biological signals monitored by \cite{li2018ictu,ciftci2019fakecatcher} by adding a discriminator which monitors these signals. To avoid anomalies in extensive the neuron activation \cite{wang2019fakespotter}, the adversary can add a loss which minimizes neuron coverage. Methods which detect abnormal poses and mannerisms \cite{agarwal2019protecting} can be evaded by reenacting the entire head and by learning the mannerisms from the same databases. Models which identify blurred content \cite{mo2018fake} are affected by noise and sharpening GANs \cite{jalalifar2018speech,kim2016accurate}, and models which search for the boundary where the face was blended in \cite{li2019face,agarwal2017swapped,zhang2017automated,akhtarcomparative,mo2018fake,durall2019unmasking} do not work on deepfakes passed through refiner networks, which use in-painting, or those which output full frames (e.g., \cite{nirkin2019fsgan,li2019faceshifter,zablotskaia2019dwnet, zhou2019dance} and many more). Finally, solutions which search for forensic evidence \cite{yu2019attributing,marra2019gans,koopman2018detection} can be evaded (or at least raise the false alarm rate) by passing the generated content through filters, or by performing physical replication or compression. 

\vspace{.5em}\noindent\textbf{Evading Deep Learning Classifiers.}
There are a number of detection methods which apply deep learning directly to the task of deepfake detection (e.g., \cite{afchar2018mesonet,do2018forensics,tariq2018detecting,ding2019swapped,fernando2019exploiting}). However, an adversary can use adversarial machine learning to evade detection by adding small perturbations to the generated image. Advances in adversarial machine learning has shown that these attacks transfer across multiple models regardless of the training data used \cite{papernot2016transferability}. Recent works have shown how these attacks not only work on deepfakes classifiers \cite{neekhara2020adversarial} but also work with no knowledge of the classifier or it's training set \cite{carlini2020evading}.

In summary, existing approaches either cannot be applied to real-time deepfakes or simply do not address a dynamic adversary which will evade their defenses.

\section{Conclusion}
Real-time deepfake attacks are starting to be used in social engineering attacks against companies and individuals. As time goes on, cyber criminals will increase their usage of this technology as a reliable means to manipulate people. Current detectors can either be evaded or cannot be used against real-time deepfakes. To counter this threat, we poposed DF-Captcha: an efficient and practical defense tool which can be used to identify deepfake callers through a challenge-response mechanism which we call a deepfake Turing test. By forcing deepfake generators to push their limitations, we can identify attacks with a much higher rate of success and with using fewer resources. More importantly, we put the defenders at an advantage since it is easier to craft challenges then it is to develop deepfake technologies.

\section*{Acknowledgments}
This work was supported by the U.S.-Israel Energy Center managed by the Israel-U.S. Binational Industrial Research and Development (BIRD) Foundation and the Zuckerman STEM Leadership Program.

\section{Bibliography}

\bibliographystyle{unsrt}
\bibliography{bibfile}

\end{document}

%% file: detection_table.tex
% \usepackage{booktabs}
% \usepackage{multirow}
% \usepackage{graphicx}
\bgroup
\def\arraystretch{.8}%  1 is the default, change whatever you need
\setlength\tabcolsep{.1em}
\begin{table*}[t]
	\centering
	\caption{Summary of Deepfake Detection Methods}

	\label{tab:detection}
	\scriptsize	
	\resizebox{.8\textwidth}{!}{%
		\begin{tabular}{@{}cccccccccccccccccccccccccc@{}}
			\toprule
			&  &  & \multicolumn{2}{c}{Type} & \multicolumn{3}{c}{Modality} & \multicolumn{4}{c}{Content} & \multicolumn{3}{c}{Method} & \multicolumn{8}{c}{Eval. Dataset} & \multicolumn{3}{c}{Performance*} \\ \midrule
			&  &  & \cellcolor[HTML]{EFEFEF}\rotatebox[origin=c]{90}{Reenactment} & \cellcolor[HTML]{EFEFEF}\rotatebox[origin=c]{90}{Replacement} & \cellcolor[HTML]{DBDBDB}\rotatebox[origin=c]{90}{Image} & \cellcolor[HTML]{DBDBDB}\rotatebox[origin=c]{90}{Video} & \cellcolor[HTML]{DBDBDB}\rotatebox[origin=c]{90}{Audio} & \cellcolor[HTML]{C0C0C0}\rotatebox[origin=c]{90}{Feature} & \cellcolor[HTML]{C0C0C0}\rotatebox[origin=c]{90}{Body Part} & \cellcolor[HTML]{C0C0C0}\rotatebox[origin=c]{90}{Face} & \cellcolor[HTML]{C0C0C0}\rotatebox[origin=c]{90}{Image} & \rotatebox[origin=c]{90}{Model} & \rotatebox[origin=c]{90}{Indicates Affected Area} & \rotatebox[origin=c]{90}{Input Resolution} & \cellcolor[HTML]{EFEFEF}\rotatebox[origin=c]{90}{DeepfakeTIMIT \cite{korshunov2018deepfakes}} & \cellcolor[HTML]{EFEFEF}\rotatebox[origin=c]{90}{DFFD \cite{stehouwer2019detection}} & \cellcolor[HTML]{EFEFEF}\rotatebox[origin=c]{90}{FaceForensics \cite{rossler2018faceforensics}} & \cellcolor[HTML]{EFEFEF}\rotatebox[origin=c]{90}{FaceForensics++ \cite{rossler2019faceforensics++}} & \cellcolor[HTML]{EFEFEF}\rotatebox[origin=c]{90}{FFW \cite{khodabakhsh2018fake}} & \cellcolor[HTML]{EFEFEF}\rotatebox[origin=c]{90}{Celeb-DF \cite{li2019celeb}} & \cellcolor[HTML]{EFEFEF}\rotatebox[origin=c]{90}{Other Deepfake DB} & \cellcolor[HTML]{EFEFEF}\rotatebox[origin=c]{90}{Custom} & \rotatebox[origin=c]{90}{ACC} & \rotatebox[origin=c]{90}{EER} & \rotatebox[origin=c]{90}{AUC} \\ \midrule\midrule
			& \cite{zhang2017automated} & 2017 & \cellcolor[HTML]{EFEFEF} & \cellcolor[HTML]{EFEFEF}$\bullet$ & \cellcolor[HTML]{DBDBDB}$\bullet$ & \cellcolor[HTML]{DBDBDB}$\bullet$ & \cellcolor[HTML]{DBDBDB} & \cellcolor[HTML]{C0C0C0} & \cellcolor[HTML]{C0C0C0} & \cellcolor[HTML]{C0C0C0}$\bullet$ & \cellcolor[HTML]{C0C0C0} & SVM-RBF &  & 250x250 & \cellcolor[HTML]{EFEFEF} & \cellcolor[HTML]{EFEFEF} & \cellcolor[HTML]{EFEFEF} & \cellcolor[HTML]{EFEFEF} & \cellcolor[HTML]{EFEFEF} & \cellcolor[HTML]{EFEFEF} & \cellcolor[HTML]{EFEFEF} & \cellcolor[HTML]{EFEFEF}$\bullet$ & 92.9 &  &  \\
			& \cite{agarwal2017swapped} & 2017 & \cellcolor[HTML]{EFEFEF} & \cellcolor[HTML]{EFEFEF}$\bullet$ & \cellcolor[HTML]{DBDBDB}$\bullet$ & \cellcolor[HTML]{DBDBDB}$\bullet$ & \cellcolor[HTML]{DBDBDB} & \cellcolor[HTML]{C0C0C0} & \cellcolor[HTML]{C0C0C0} & \cellcolor[HTML]{C0C0C0} & \cellcolor[HTML]{C0C0C0}$\bullet$ & SVM &  & * & \cellcolor[HTML]{EFEFEF} & \cellcolor[HTML]{EFEFEF} & \cellcolor[HTML]{EFEFEF} & \cellcolor[HTML]{EFEFEF} & \cellcolor[HTML]{EFEFEF} & \cellcolor[HTML]{EFEFEF} & \cellcolor[HTML]{EFEFEF}$\bullet$ & \cellcolor[HTML]{EFEFEF} &  & 18.2 &  \\
			& \cite{yang2019exposing} & 2018 & \cellcolor[HTML]{EFEFEF}$\bullet$ & \cellcolor[HTML]{EFEFEF}$\bullet$ & \cellcolor[HTML]{DBDBDB}$\bullet$ & \cellcolor[HTML]{DBDBDB}$\bullet$ & \cellcolor[HTML]{DBDBDB} & \cellcolor[HTML]{C0C0C0} & \cellcolor[HTML]{C0C0C0} & \cellcolor[HTML]{C0C0C0}$\bullet$ & \cellcolor[HTML]{C0C0C0} & SVM &  & * & \cellcolor[HTML]{EFEFEF} & \cellcolor[HTML]{EFEFEF} & \cellcolor[HTML]{EFEFEF} & \cellcolor[HTML]{EFEFEF} & \cellcolor[HTML]{EFEFEF} & \cellcolor[HTML]{EFEFEF} & \cellcolor[HTML]{EFEFEF}$\bullet$ & \cellcolor[HTML]{EFEFEF} &  &  & 0.97 \\
			& \cite{korshunov2018deepfakes} & 2018 & \cellcolor[HTML]{EFEFEF}$\bullet$ & \cellcolor[HTML]{EFEFEF}$\bullet$ & \cellcolor[HTML]{DBDBDB} & \cellcolor[HTML]{DBDBDB}$\bullet$ & \cellcolor[HTML]{DBDBDB} & \cellcolor[HTML]{C0C0C0} & \cellcolor[HTML]{C0C0C0} & \cellcolor[HTML]{C0C0C0}$\bullet$ & \cellcolor[HTML]{C0C0C0} & SVM &  & 128x128 & \cellcolor[HTML]{EFEFEF}$\bullet$ & \cellcolor[HTML]{EFEFEF} & \cellcolor[HTML]{EFEFEF} & \cellcolor[HTML]{EFEFEF} & \cellcolor[HTML]{EFEFEF} & \cellcolor[HTML]{EFEFEF} & \cellcolor[HTML]{EFEFEF} & \cellcolor[HTML]{EFEFEF} &  & 3.33 &  \\
			& \cite{durall2019unmasking} & 2019 & \cellcolor[HTML]{EFEFEF} & \cellcolor[HTML]{EFEFEF}$\bullet$ & \cellcolor[HTML]{DBDBDB}$\bullet$ & \cellcolor[HTML]{DBDBDB} & \cellcolor[HTML]{DBDBDB} & \cellcolor[HTML]{C0C0C0} & \cellcolor[HTML]{C0C0C0} & \cellcolor[HTML]{C0C0C0} & \cellcolor[HTML]{C0C0C0}$\bullet$ & SVM, Kmeans... &  & 1024x1024 & \cellcolor[HTML]{EFEFEF} & \cellcolor[HTML]{EFEFEF} & \cellcolor[HTML]{EFEFEF} & \cellcolor[HTML]{EFEFEF} & \cellcolor[HTML]{EFEFEF} & \cellcolor[HTML]{EFEFEF} & \cellcolor[HTML]{EFEFEF} & \cellcolor[HTML]{EFEFEF}$\bullet$ & 100 &  &  \\
			& \cite{akhtarcomparative} & 2019 & \cellcolor[HTML]{EFEFEF}$\bullet$ & \cellcolor[HTML]{EFEFEF}$\bullet$ & \cellcolor[HTML]{DBDBDB}$\bullet$ & \cellcolor[HTML]{DBDBDB}$\bullet$ & \cellcolor[HTML]{DBDBDB} & \cellcolor[HTML]{C0C0C0} & \cellcolor[HTML]{C0C0C0} & \cellcolor[HTML]{C0C0C0}$\bullet$ & \cellcolor[HTML]{C0C0C0} & SVM &  & * & \cellcolor[HTML]{EFEFEF}$\bullet$ & \cellcolor[HTML]{EFEFEF} & \cellcolor[HTML]{EFEFEF} & \cellcolor[HTML]{EFEFEF} & \cellcolor[HTML]{EFEFEF} & \cellcolor[HTML]{EFEFEF} & \cellcolor[HTML]{EFEFEF} & \cellcolor[HTML]{EFEFEF} &  & 13.33 &  \\
			\multirow{-7}{*}{Classic ML} & \cite{agarwal2019protecting} & 2019 & \cellcolor[HTML]{EFEFEF}$\bullet$ & \cellcolor[HTML]{EFEFEF}$\bullet$ & \cellcolor[HTML]{DBDBDB} & \cellcolor[HTML]{DBDBDB}$\bullet$ & \cellcolor[HTML]{DBDBDB} & \cellcolor[HTML]{C0C0C0} & \cellcolor[HTML]{C0C0C0} & \cellcolor[HTML]{C0C0C0}$\bullet$ & \cellcolor[HTML]{C0C0C0} & SVM &  & * & \cellcolor[HTML]{EFEFEF} & \cellcolor[HTML]{EFEFEF} & \cellcolor[HTML]{EFEFEF} & \cellcolor[HTML]{EFEFEF} & \cellcolor[HTML]{EFEFEF} & \cellcolor[HTML]{EFEFEF} & \cellcolor[HTML]{EFEFEF} & \cellcolor[HTML]{EFEFEF}$\bullet$ &  &  & 0.98 \\ \midrule
			& \cite{mo2018fake} & 2018 & \cellcolor[HTML]{EFEFEF}$\bullet$ & \cellcolor[HTML]{EFEFEF}$\bullet$ & \cellcolor[HTML]{DBDBDB}$\bullet$ & \cellcolor[HTML]{DBDBDB}$\bullet$ & \cellcolor[HTML]{DBDBDB} & \cellcolor[HTML]{C0C0C0} & \cellcolor[HTML]{C0C0C0} & \cellcolor[HTML]{C0C0C0}$\bullet$ & \cellcolor[HTML]{C0C0C0} & CNN &  & 256x256 & \cellcolor[HTML]{EFEFEF} & \cellcolor[HTML]{EFEFEF} & \cellcolor[HTML]{EFEFEF} & \cellcolor[HTML]{EFEFEF} & \cellcolor[HTML]{EFEFEF} & \cellcolor[HTML]{EFEFEF} & \cellcolor[HTML]{EFEFEF} & \cellcolor[HTML]{EFEFEF}$\bullet$ & 99.4 &  &  \\
			& \cite{li2018ictu} & 2018 & \cellcolor[HTML]{EFEFEF}$\bullet$ & \cellcolor[HTML]{EFEFEF}$\bullet$ & \cellcolor[HTML]{DBDBDB} & \cellcolor[HTML]{DBDBDB} & \cellcolor[HTML]{DBDBDB} & \cellcolor[HTML]{C0C0C0}$\bullet$ & \cellcolor[HTML]{C0C0C0}$\bullet$ & \cellcolor[HTML]{C0C0C0} & \cellcolor[HTML]{C0C0C0} & LSTM-CNN &  & 224x224 & \cellcolor[HTML]{EFEFEF} & \cellcolor[HTML]{EFEFEF} & \cellcolor[HTML]{EFEFEF} & \cellcolor[HTML]{EFEFEF} & \cellcolor[HTML]{EFEFEF} & \cellcolor[HTML]{EFEFEF} & \cellcolor[HTML]{EFEFEF} & \cellcolor[HTML]{EFEFEF}$\bullet$ &  &  & 0.99 \\
			& \cite{nguyen2019capsule} & 2018 & \cellcolor[HTML]{EFEFEF}$\bullet$ & \cellcolor[HTML]{EFEFEF}$\bullet$ & \cellcolor[HTML]{DBDBDB}$\bullet$ & \cellcolor[HTML]{DBDBDB}$\bullet$ & \cellcolor[HTML]{DBDBDB} & \cellcolor[HTML]{C0C0C0} & \cellcolor[HTML]{C0C0C0} & \cellcolor[HTML]{C0C0C0}$\bullet$ & \cellcolor[HTML]{C0C0C0} & Capsule-CNN &  & 128x128 & \cellcolor[HTML]{EFEFEF} & \cellcolor[HTML]{EFEFEF} & \cellcolor[HTML]{EFEFEF}$\bullet$ & \cellcolor[HTML]{EFEFEF} & \cellcolor[HTML]{EFEFEF} & \cellcolor[HTML]{EFEFEF} & \cellcolor[HTML]{EFEFEF} & \cellcolor[HTML]{EFEFEF} & 99.3 &  &  \\
			& \cite{bao2018towards} & 2018 & \cellcolor[HTML]{EFEFEF} & \cellcolor[HTML]{EFEFEF}$\bullet$ & \cellcolor[HTML]{DBDBDB}$\bullet$ & \cellcolor[HTML]{DBDBDB}$\bullet$ & \cellcolor[HTML]{DBDBDB} & \cellcolor[HTML]{C0C0C0} & \cellcolor[HTML]{C0C0C0} & \cellcolor[HTML]{C0C0C0}$\bullet$ & \cellcolor[HTML]{C0C0C0} & ED-GAN &  & 128x128 & \cellcolor[HTML]{EFEFEF} & \cellcolor[HTML]{EFEFEF} & \cellcolor[HTML]{EFEFEF} & \cellcolor[HTML]{EFEFEF} & \cellcolor[HTML]{EFEFEF} & \cellcolor[HTML]{EFEFEF} & \cellcolor[HTML]{EFEFEF} & \cellcolor[HTML]{EFEFEF}$\bullet$ & 92 &  &  \\
			& \cite{do2018forensics} & 2018 & \cellcolor[HTML]{EFEFEF}$\bullet$ & \cellcolor[HTML]{EFEFEF}$\bullet$ & \cellcolor[HTML]{DBDBDB}$\bullet$ & \cellcolor[HTML]{DBDBDB} & \cellcolor[HTML]{DBDBDB} & \cellcolor[HTML]{C0C0C0} & \cellcolor[HTML]{C0C0C0} & \cellcolor[HTML]{C0C0C0}$\bullet$ & \cellcolor[HTML]{C0C0C0} & CNN &  & 1024x1024 & \cellcolor[HTML]{EFEFEF} & \cellcolor[HTML]{EFEFEF} & \cellcolor[HTML]{EFEFEF} & \cellcolor[HTML]{EFEFEF} & \cellcolor[HTML]{EFEFEF} & \cellcolor[HTML]{EFEFEF} & \cellcolor[HTML]{EFEFEF}$\bullet$ & \cellcolor[HTML]{EFEFEF} &  &  & 0.81 \\
			& \cite{guera2018deepfake} & 2018 & \cellcolor[HTML]{EFEFEF}$\bullet$ & \cellcolor[HTML]{EFEFEF}$\bullet$ & \cellcolor[HTML]{DBDBDB} & \cellcolor[HTML]{DBDBDB}$\bullet$ & \cellcolor[HTML]{DBDBDB} & \cellcolor[HTML]{C0C0C0} & \cellcolor[HTML]{C0C0C0} & \cellcolor[HTML]{C0C0C0}$\bullet$ & \cellcolor[HTML]{C0C0C0} & CNN-LSTM &  & 299x299 & \cellcolor[HTML]{EFEFEF} & \cellcolor[HTML]{EFEFEF} & \cellcolor[HTML]{EFEFEF} & \cellcolor[HTML]{EFEFEF} & \cellcolor[HTML]{EFEFEF} & \cellcolor[HTML]{EFEFEF} & \cellcolor[HTML]{EFEFEF} & \cellcolor[HTML]{EFEFEF}$\bullet$ & 97.1 &  &  \\
			& \cite{marra2018detection} & 2018 & \cellcolor[HTML]{EFEFEF}$\bullet$ & \cellcolor[HTML]{EFEFEF}$\bullet$ & \cellcolor[HTML]{DBDBDB}$\bullet$ & \cellcolor[HTML]{DBDBDB} & \cellcolor[HTML]{DBDBDB} & \cellcolor[HTML]{C0C0C0} & \cellcolor[HTML]{C0C0C0} & \cellcolor[HTML]{C0C0C0} & \cellcolor[HTML]{C0C0C0}$\bullet$ & CNN &  & 256x256 & \cellcolor[HTML]{EFEFEF} & \cellcolor[HTML]{EFEFEF} & \cellcolor[HTML]{EFEFEF} & \cellcolor[HTML]{EFEFEF} & \cellcolor[HTML]{EFEFEF} & \cellcolor[HTML]{EFEFEF} & \cellcolor[HTML]{EFEFEF} & \cellcolor[HTML]{EFEFEF}$\bullet$ & 94.4 &  &  \\
			& \cite{cozzolino2018forensictransfer} & 2018 & \cellcolor[HTML]{EFEFEF}$\bullet$ & \cellcolor[HTML]{EFEFEF}$\bullet$ & \cellcolor[HTML]{DBDBDB}$\bullet$ & \cellcolor[HTML]{DBDBDB}$\bullet$ & \cellcolor[HTML]{DBDBDB} & \cellcolor[HTML]{C0C0C0} & \cellcolor[HTML]{C0C0C0} & \cellcolor[HTML]{C0C0C0} & \cellcolor[HTML]{C0C0C0}$\bullet$ & CNN AE &  & 256x256 & \cellcolor[HTML]{EFEFEF} & \cellcolor[HTML]{EFEFEF} & \cellcolor[HTML]{EFEFEF}$\bullet$ & \cellcolor[HTML]{EFEFEF} & \cellcolor[HTML]{EFEFEF} & \cellcolor[HTML]{EFEFEF} & \cellcolor[HTML]{EFEFEF} & \cellcolor[HTML]{EFEFEF}$\bullet$ & 90.5 &  &  \\
			& \cite{afchar2018mesonet} & 2018 & \cellcolor[HTML]{EFEFEF}$\bullet$ & \cellcolor[HTML]{EFEFEF}$\bullet$ & \cellcolor[HTML]{DBDBDB} & \cellcolor[HTML]{DBDBDB}$\bullet$ & \cellcolor[HTML]{DBDBDB} & \cellcolor[HTML]{C0C0C0} & \cellcolor[HTML]{C0C0C0} & \cellcolor[HTML]{C0C0C0}$\bullet$ & \cellcolor[HTML]{C0C0C0} & CNN &  & 256x256 & \cellcolor[HTML]{EFEFEF} & \cellcolor[HTML]{EFEFEF} & \cellcolor[HTML]{EFEFEF} & \cellcolor[HTML]{EFEFEF} & \cellcolor[HTML]{EFEFEF} & \cellcolor[HTML]{EFEFEF} & \cellcolor[HTML]{EFEFEF} & \cellcolor[HTML]{EFEFEF}$\bullet$ &  &  & 0.99 \\
			& \cite{sabir2019recurrent} & 2019 & \cellcolor[HTML]{EFEFEF}$\bullet$ & \cellcolor[HTML]{EFEFEF}$\bullet$ & \cellcolor[HTML]{DBDBDB} & \cellcolor[HTML]{DBDBDB}$\bullet$ & \cellcolor[HTML]{DBDBDB} & \cellcolor[HTML]{C0C0C0} & \cellcolor[HTML]{C0C0C0} & \cellcolor[HTML]{C0C0C0}$\bullet$ & \cellcolor[HTML]{C0C0C0} & CNN-LSTM &  & 224x224 & \cellcolor[HTML]{EFEFEF} & \cellcolor[HTML]{EFEFEF} & \cellcolor[HTML]{EFEFEF} & \cellcolor[HTML]{EFEFEF}$\bullet$ & \cellcolor[HTML]{EFEFEF} & \cellcolor[HTML]{EFEFEF} & \cellcolor[HTML]{EFEFEF} & \cellcolor[HTML]{EFEFEF} & 96.9 &  &  \\
			& \cite{nguyen2019multi} & 2019 & \cellcolor[HTML]{EFEFEF}$\bullet$ & \cellcolor[HTML]{EFEFEF}$\bullet$ & \cellcolor[HTML]{DBDBDB}$\bullet$ & \cellcolor[HTML]{DBDBDB}$\bullet$ & \cellcolor[HTML]{DBDBDB} & \cellcolor[HTML]{C0C0C0} & \cellcolor[HTML]{C0C0C0} & \cellcolor[HTML]{C0C0C0}$\bullet$ & \cellcolor[HTML]{C0C0C0} & CNN-DE & $\bullet$ & 256x256 & \cellcolor[HTML]{EFEFEF} & \cellcolor[HTML]{EFEFEF} & \cellcolor[HTML]{EFEFEF}$\bullet$ & \cellcolor[HTML]{EFEFEF}$\bullet$ & \cellcolor[HTML]{EFEFEF} & \cellcolor[HTML]{EFEFEF} & \cellcolor[HTML]{EFEFEF} & \cellcolor[HTML]{EFEFEF} & 92.8 & 8.18 &  \\
			& \cite{ding2019swapped} & 2019 & \cellcolor[HTML]{EFEFEF} & \cellcolor[HTML]{EFEFEF}$\bullet$ & \cellcolor[HTML]{DBDBDB}$\bullet$ & \cellcolor[HTML]{DBDBDB}$\bullet$ & \cellcolor[HTML]{DBDBDB} & \cellcolor[HTML]{C0C0C0} & \cellcolor[HTML]{C0C0C0} & \cellcolor[HTML]{C0C0C0}$\bullet$ & \cellcolor[HTML]{C0C0C0} & CNN &  & - & \cellcolor[HTML]{EFEFEF} & \cellcolor[HTML]{EFEFEF} & \cellcolor[HTML]{EFEFEF} & \cellcolor[HTML]{EFEFEF} & \cellcolor[HTML]{EFEFEF} & \cellcolor[HTML]{EFEFEF} & \cellcolor[HTML]{EFEFEF} & \cellcolor[HTML]{EFEFEF}$\bullet$ & 98.5 &  &  \\
			& \cite{du2019towards} & 2019 & \cellcolor[HTML]{EFEFEF}$\bullet$ & \cellcolor[HTML]{EFEFEF}$\bullet$ & \cellcolor[HTML]{DBDBDB}$\bullet$ & \cellcolor[HTML]{DBDBDB}$\bullet$ & \cellcolor[HTML]{DBDBDB} & \cellcolor[HTML]{C0C0C0} & \cellcolor[HTML]{C0C0C0} & \cellcolor[HTML]{C0C0C0} & \cellcolor[HTML]{C0C0C0}$\bullet$ & CNN AE GAN & $\bullet$ & 256x256 & \cellcolor[HTML]{EFEFEF} & \cellcolor[HTML]{EFEFEF} & \cellcolor[HTML]{EFEFEF} & \cellcolor[HTML]{EFEFEF}$\bullet$ & \cellcolor[HTML]{EFEFEF} & \cellcolor[HTML]{EFEFEF} & \cellcolor[HTML]{EFEFEF} & \cellcolor[HTML]{EFEFEF}$\bullet$ & 99.2 &  &  \\
			& \cite{stehouwer2019detection} & 2019 & \cellcolor[HTML]{EFEFEF}$\bullet$ & \cellcolor[HTML]{EFEFEF}$\bullet$ & \cellcolor[HTML]{DBDBDB}$\bullet$ & \cellcolor[HTML]{DBDBDB}$\bullet$ & \cellcolor[HTML]{DBDBDB} & \cellcolor[HTML]{C0C0C0} & \cellcolor[HTML]{C0C0C0} & \cellcolor[HTML]{C0C0C0}$\bullet$ & \cellcolor[HTML]{C0C0C0} & CNN+Attention & $\bullet$ & 299x299 & \cellcolor[HTML]{EFEFEF} & \cellcolor[HTML]{EFEFEF}$\bullet$ & \cellcolor[HTML]{EFEFEF} & \cellcolor[HTML]{EFEFEF} & \cellcolor[HTML]{EFEFEF} & \cellcolor[HTML]{EFEFEF} & \cellcolor[HTML]{EFEFEF} & \cellcolor[HTML]{EFEFEF} &  & 3.11 & 0.99 \\
			& \cite{danmohah54:online} & 2019 & \cellcolor[HTML]{EFEFEF}$\bullet$ & \cellcolor[HTML]{EFEFEF}$\bullet$ & \cellcolor[HTML]{DBDBDB}$\bullet$ & \cellcolor[HTML]{DBDBDB}$\bullet$ & \cellcolor[HTML]{DBDBDB} & \cellcolor[HTML]{C0C0C0} & \cellcolor[HTML]{C0C0C0} & \cellcolor[HTML]{C0C0C0}$\bullet$ & \cellcolor[HTML]{C0C0C0} & CNN &  & 128x128 & \cellcolor[HTML]{EFEFEF} & \cellcolor[HTML]{EFEFEF} & \cellcolor[HTML]{EFEFEF} & \cellcolor[HTML]{EFEFEF} & \cellcolor[HTML]{EFEFEF} & \cellcolor[HTML]{EFEFEF} & \cellcolor[HTML]{EFEFEF}$\bullet$ & \cellcolor[HTML]{EFEFEF}$\bullet$ &  &  & 0.99 \\
			& \cite{li2019celeb} & 2019 & \cellcolor[HTML]{EFEFEF} & \cellcolor[HTML]{EFEFEF}$\bullet$ & \cellcolor[HTML]{DBDBDB} & \cellcolor[HTML]{DBDBDB}$\bullet$ & \cellcolor[HTML]{DBDBDB} & \cellcolor[HTML]{C0C0C0} & \cellcolor[HTML]{C0C0C0} & \cellcolor[HTML]{C0C0C0}$\bullet$ & \cellcolor[HTML]{C0C0C0} & CNN &  & * & \cellcolor[HTML]{EFEFEF} & \cellcolor[HTML]{EFEFEF} & \cellcolor[HTML]{EFEFEF} & \cellcolor[HTML]{EFEFEF} & \cellcolor[HTML]{EFEFEF} & \cellcolor[HTML]{EFEFEF}$\bullet$ & \cellcolor[HTML]{EFEFEF} & \cellcolor[HTML]{EFEFEF} &  &  & 0.64 \\
			& \cite{fernando2019exploiting} & 2019 & \cellcolor[HTML]{EFEFEF}$\bullet$ & \cellcolor[HTML]{EFEFEF}$\bullet$ & \cellcolor[HTML]{DBDBDB}$\bullet$ & \cellcolor[HTML]{DBDBDB}$\bullet$ & \cellcolor[HTML]{DBDBDB} & \cellcolor[HTML]{C0C0C0} & \cellcolor[HTML]{C0C0C0} & \cellcolor[HTML]{C0C0C0}$\bullet$ & \cellcolor[HTML]{C0C0C0} & CNN+HMN &  & 224x224 & \cellcolor[HTML]{EFEFEF} & \cellcolor[HTML]{EFEFEF} & \cellcolor[HTML]{EFEFEF}$\bullet$ & \cellcolor[HTML]{EFEFEF}$\bullet$ & \cellcolor[HTML]{EFEFEF}$\bullet$ & \cellcolor[HTML]{EFEFEF} & \cellcolor[HTML]{EFEFEF} & \cellcolor[HTML]{EFEFEF} & 99.4 &  &  \\
			& \cite{li2019zooming} & 2019 & \cellcolor[HTML]{EFEFEF}$\bullet$ & \cellcolor[HTML]{EFEFEF}$\bullet$ & \cellcolor[HTML]{DBDBDB}$\bullet$ & \cellcolor[HTML]{DBDBDB}$\bullet$ & \cellcolor[HTML]{DBDBDB} & \cellcolor[HTML]{C0C0C0} & \cellcolor[HTML]{C0C0C0} & \cellcolor[HTML]{C0C0C0}$\bullet$ & \cellcolor[HTML]{C0C0C0} & FCN &  & 256x256 & \cellcolor[HTML]{EFEFEF} & \cellcolor[HTML]{EFEFEF} & \cellcolor[HTML]{EFEFEF} & \cellcolor[HTML]{EFEFEF}$\bullet$ & \cellcolor[HTML]{EFEFEF} & \cellcolor[HTML]{EFEFEF} & \cellcolor[HTML]{EFEFEF} & \cellcolor[HTML]{EFEFEF} & 98.1 &  &  \\
			& \cite{xuan2019generalization} & 2019 & \cellcolor[HTML]{EFEFEF}$\bullet$ & \cellcolor[HTML]{EFEFEF}$\bullet$ & \cellcolor[HTML]{DBDBDB}$\bullet$ & \cellcolor[HTML]{DBDBDB}$\bullet$ & \cellcolor[HTML]{DBDBDB} & \cellcolor[HTML]{C0C0C0} & \cellcolor[HTML]{C0C0C0} & \cellcolor[HTML]{C0C0C0} & \cellcolor[HTML]{C0C0C0}$\bullet$ & CNN &  & 128x128 & \cellcolor[HTML]{EFEFEF} & \cellcolor[HTML]{EFEFEF} & \cellcolor[HTML]{EFEFEF} & \cellcolor[HTML]{EFEFEF} & \cellcolor[HTML]{EFEFEF} & \cellcolor[HTML]{EFEFEF} & \cellcolor[HTML]{EFEFEF} & \cellcolor[HTML]{EFEFEF}$\bullet$ & 94.7 &  &  \\
			& \cite{tu2019deep} & 2019 & \cellcolor[HTML]{EFEFEF}$\bullet$ & \cellcolor[HTML]{EFEFEF}$\bullet$ & \cellcolor[HTML]{DBDBDB}$\bullet$ & \cellcolor[HTML]{DBDBDB}$\bullet$ & \cellcolor[HTML]{DBDBDB} & \cellcolor[HTML]{C0C0C0} & \cellcolor[HTML]{C0C0C0} & \cellcolor[HTML]{C0C0C0}$\bullet$ & \cellcolor[HTML]{C0C0C0} & CNN &  & 224x224 & \cellcolor[HTML]{EFEFEF} & \cellcolor[HTML]{EFEFEF} & \cellcolor[HTML]{EFEFEF} & \cellcolor[HTML]{EFEFEF} & \cellcolor[HTML]{EFEFEF} & \cellcolor[HTML]{EFEFEF} & \cellcolor[HTML]{EFEFEF}$\bullet$ & \cellcolor[HTML]{EFEFEF} & 86.4 &  &  \\
			& \cite{tariq2018detecting} & 2019 & \cellcolor[HTML]{EFEFEF}$\bullet$ & \cellcolor[HTML]{EFEFEF}$\bullet$ & \cellcolor[HTML]{DBDBDB}$\bullet$ & \cellcolor[HTML]{DBDBDB} & \cellcolor[HTML]{DBDBDB} & \cellcolor[HTML]{C0C0C0} & \cellcolor[HTML]{C0C0C0} & \cellcolor[HTML]{C0C0C0}$\bullet$ & \cellcolor[HTML]{C0C0C0} & CNN &  & 1024x1024 & \cellcolor[HTML]{EFEFEF} & \cellcolor[HTML]{EFEFEF} & \cellcolor[HTML]{EFEFEF} & \cellcolor[HTML]{EFEFEF} & \cellcolor[HTML]{EFEFEF} & \cellcolor[HTML]{EFEFEF} & \cellcolor[HTML]{EFEFEF} & \cellcolor[HTML]{EFEFEF}$\bullet$ &  &  & 94 \\
			& \cite{ciftci2019fakecatcher} & 2019 & \cellcolor[HTML]{EFEFEF}$\bullet$ & \cellcolor[HTML]{EFEFEF}$\bullet$ & \cellcolor[HTML]{DBDBDB} & \cellcolor[HTML]{DBDBDB}$\bullet$ & \cellcolor[HTML]{DBDBDB} & \cellcolor[HTML]{C0C0C0} & \cellcolor[HTML]{C0C0C0} & \cellcolor[HTML]{C0C0C0}$\bullet$ & \cellcolor[HTML]{C0C0C0} & CNN &  & 128x128 & \cellcolor[HTML]{EFEFEF} & \cellcolor[HTML]{EFEFEF} & \cellcolor[HTML]{EFEFEF}$\bullet$ & \cellcolor[HTML]{EFEFEF} & \cellcolor[HTML]{EFEFEF} & \cellcolor[HTML]{EFEFEF} & \cellcolor[HTML]{EFEFEF} & \cellcolor[HTML]{EFEFEF}$\bullet$ & 96 &  &  \\
			& \cite{li2019exposing} & 2019 & \cellcolor[HTML]{EFEFEF}$\bullet$ & \cellcolor[HTML]{EFEFEF}$\bullet$ & \cellcolor[HTML]{DBDBDB}$\bullet$ & \cellcolor[HTML]{DBDBDB}$\bullet$ & \cellcolor[HTML]{DBDBDB} & \cellcolor[HTML]{C0C0C0} & \cellcolor[HTML]{C0C0C0} & \cellcolor[HTML]{C0C0C0} & \cellcolor[HTML]{C0C0C0}$\bullet$ & CNN & $\bullet$ & 224x224 & \cellcolor[HTML]{EFEFEF}$\bullet$ & \cellcolor[HTML]{EFEFEF} & \cellcolor[HTML]{EFEFEF} & \cellcolor[HTML]{EFEFEF} & \cellcolor[HTML]{EFEFEF} & \cellcolor[HTML]{EFEFEF} & \cellcolor[HTML]{EFEFEF} & \cellcolor[HTML]{EFEFEF} &  & 93.2 &  \\
			& \cite{amerini2019deepfake} & 2019 & \cellcolor[HTML]{EFEFEF}$\bullet$ & \cellcolor[HTML]{EFEFEF}$\bullet$ & \cellcolor[HTML]{DBDBDB} & \cellcolor[HTML]{DBDBDB}$\bullet$ & \cellcolor[HTML]{DBDBDB} & \cellcolor[HTML]{C0C0C0} & \cellcolor[HTML]{C0C0C0} & \cellcolor[HTML]{C0C0C0}$\bullet$ & \cellcolor[HTML]{C0C0C0} & CNN &  & 224x224 & \cellcolor[HTML]{EFEFEF} & \cellcolor[HTML]{EFEFEF} & \cellcolor[HTML]{EFEFEF} & \cellcolor[HTML]{EFEFEF}$\bullet$ & \cellcolor[HTML]{EFEFEF} & \cellcolor[HTML]{EFEFEF} & \cellcolor[HTML]{EFEFEF} & \cellcolor[HTML]{EFEFEF} & 81.6 &  &  \\
			& \cite{korshunovspeaker} & 2019 & \cellcolor[HTML]{EFEFEF}$\bullet$ & \cellcolor[HTML]{EFEFEF} & \cellcolor[HTML]{DBDBDB} & \cellcolor[HTML]{DBDBDB}$\bullet$ & \cellcolor[HTML]{DBDBDB}$\bullet$ & \cellcolor[HTML]{C0C0C0}$\bullet$ & \cellcolor[HTML]{C0C0C0}$\bullet$ & \cellcolor[HTML]{C0C0C0} & \cellcolor[HTML]{C0C0C0} & LSTM &  & * & \cellcolor[HTML]{EFEFEF} & \cellcolor[HTML]{EFEFEF} & \cellcolor[HTML]{EFEFEF} & \cellcolor[HTML]{EFEFEF} & \cellcolor[HTML]{EFEFEF} & \cellcolor[HTML]{EFEFEF} & \cellcolor[HTML]{EFEFEF} & \cellcolor[HTML]{EFEFEF}$\bullet$ &  & 22 &  \\
			& \cite{korshunov2019tampered} & 2019 & \cellcolor[HTML]{EFEFEF}$\bullet$ & \cellcolor[HTML]{EFEFEF} & \cellcolor[HTML]{DBDBDB} & \cellcolor[HTML]{DBDBDB}$\bullet$ & \cellcolor[HTML]{DBDBDB}$\bullet$ & \cellcolor[HTML]{C0C0C0}$\bullet$ & \cellcolor[HTML]{C0C0C0}$\bullet$ & \cellcolor[HTML]{C0C0C0} & \cellcolor[HTML]{C0C0C0} & LSTM-DNN &  & * & \cellcolor[HTML]{EFEFEF} & \cellcolor[HTML]{EFEFEF} & \cellcolor[HTML]{EFEFEF} & \cellcolor[HTML]{EFEFEF} & \cellcolor[HTML]{EFEFEF} & \cellcolor[HTML]{EFEFEF} & \cellcolor[HTML]{EFEFEF} & \cellcolor[HTML]{EFEFEF}$\bullet$ &  & 16.4 &  \\
			& \cite{chan2019everybody} & 2019 & \cellcolor[HTML]{EFEFEF}$\bullet$ & \cellcolor[HTML]{EFEFEF} & \cellcolor[HTML]{DBDBDB} & \cellcolor[HTML]{DBDBDB}$\bullet$ & \cellcolor[HTML]{DBDBDB} & \cellcolor[HTML]{C0C0C0} & \cellcolor[HTML]{C0C0C0} & \cellcolor[HTML]{C0C0C0} & \cellcolor[HTML]{C0C0C0}$\bullet$ & CNN &  & 256x256 & \cellcolor[HTML]{EFEFEF} & \cellcolor[HTML]{EFEFEF} & \cellcolor[HTML]{EFEFEF} & \cellcolor[HTML]{EFEFEF} & \cellcolor[HTML]{EFEFEF} & \cellcolor[HTML]{EFEFEF} & \cellcolor[HTML]{EFEFEF} & \cellcolor[HTML]{EFEFEF}$\bullet$ & 97 &  &  \\
			\multirow{-29}{*}{Deep Learning} & \cite{yu2019attributing} & 2019 & \cellcolor[HTML]{EFEFEF}$\bullet$ & \cellcolor[HTML]{EFEFEF}$\bullet$ & \cellcolor[HTML]{DBDBDB}$\bullet$ & \cellcolor[HTML]{DBDBDB} & \cellcolor[HTML]{DBDBDB} & \cellcolor[HTML]{C0C0C0}$\bullet$ & \cellcolor[HTML]{C0C0C0} & \cellcolor[HTML]{C0C0C0} & \cellcolor[HTML]{C0C0C0}$\bullet$ & CNN &  & 128x128 & \cellcolor[HTML]{EFEFEF} & \cellcolor[HTML]{EFEFEF} & \cellcolor[HTML]{EFEFEF}$\bullet$ & \cellcolor[HTML]{EFEFEF} & \cellcolor[HTML]{EFEFEF}$\bullet$ & \cellcolor[HTML]{EFEFEF} & \cellcolor[HTML]{EFEFEF} & \cellcolor[HTML]{EFEFEF}$\bullet$ & 99.6 & 0.53 &  \\ 
			& \cite{wang2019fakespotter} & 2019 & \cellcolor[HTML]{EFEFEF}$\bullet$ & \cellcolor[HTML]{EFEFEF}$\bullet$ & \cellcolor[HTML]{DBDBDB}$\bullet$ & \cellcolor[HTML]{DBDBDB}$\bullet$ & \cellcolor[HTML]{DBDBDB} & \cellcolor[HTML]{C0C0C0} & \cellcolor[HTML]{C0C0C0} & \cellcolor[HTML]{C0C0C0}$\bullet$ & \cellcolor[HTML]{C0C0C0} & SVM+VGGnet &  & 224x224 & \cellcolor[HTML]{EFEFEF} & \cellcolor[HTML]{EFEFEF} & \cellcolor[HTML]{EFEFEF} & \cellcolor[HTML]{EFEFEF}$\bullet$ & \cellcolor[HTML]{EFEFEF} & \cellcolor[HTML]{EFEFEF} & \cellcolor[HTML]{EFEFEF} & \cellcolor[HTML]{EFEFEF} & 85 &  &  \\ 
			& \cite{li2019face} & 2019 & \cellcolor[HTML]{EFEFEF}$\bullet$ & \cellcolor[HTML]{EFEFEF}$\bullet$ & \cellcolor[HTML]{DBDBDB}$\bullet$ & \cellcolor[HTML]{DBDBDB}$\bullet$ & \cellcolor[HTML]{DBDBDB} & \cellcolor[HTML]{C0C0C0}$\bullet$ & \cellcolor[HTML]{C0C0C0} & \cellcolor[HTML]{C0C0C0} & \cellcolor[HTML]{C0C0C0} & CNN & $\bullet$ & 64x64 & \cellcolor[HTML]{EFEFEF} & \cellcolor[HTML]{EFEFEF}$\bullet$ & \cellcolor[HTML]{EFEFEF} & \cellcolor[HTML]{EFEFEF}$\bullet$ & \cellcolor[HTML]{EFEFEF} & \cellcolor[HTML]{EFEFEF}$\bullet$ & \cellcolor[HTML]{EFEFEF}$\bullet$ & \cellcolor[HTML]{EFEFEF} &  &  & 99.2 \\ 
			
			&\cite{li2020face}  & 2020 &  \cellcolor[HTML]{EFEFEF}$\circ$  & \cellcolor[HTML]{EFEFEF} $\bullet$  & \cellcolor[HTML]{DBDBDB} $\bullet$  & \cellcolor[HTML]{DBDBDB}   & \cellcolor[HTML]{DBDBDB}   & \cellcolor[HTML]{C0C0C0}   & \cellcolor[HTML]{C0C0C0}   & \cellcolor[HTML]{C0C0C0} $\bullet$  & \cellcolor[HTML]{C0C0C0} $\bullet$  &  HRNet-FCN  &  $\bullet$  & 64x64  & \cellcolor[HTML]{EFEFEF}   & \cellcolor[HTML]{EFEFEF} $\bullet$  & \cellcolor[HTML]{EFEFEF}   & \cellcolor[HTML]{EFEFEF} $\bullet$  & \cellcolor[HTML]{EFEFEF}   & \cellcolor[HTML]{EFEFEF}   & \cellcolor[HTML]{EFEFEF} $\bullet$  & \cellcolor[HTML]{EFEFEF}   &    & 20.86 &  0.86\\
			&\cite{li2020fighting}  & 2020 &  \cellcolor[HTML]{EFEFEF}$\bullet$  & \cellcolor[HTML]{EFEFEF} $\bullet$  & \cellcolor[HTML]{DBDBDB} $\bullet$  & \cellcolor[HTML]{DBDBDB}   & \cellcolor[HTML]{DBDBDB}   & \cellcolor[HTML]{C0C0C0}   & \cellcolor[HTML]{C0C0C0}   & \cellcolor[HTML]{C0C0C0} $\bullet$  & \cellcolor[HTML]{C0C0C0} $\bullet$  &  PP-CNN  &    &  -  & \cellcolor[HTML]{EFEFEF} $\bullet$  & \cellcolor[HTML]{EFEFEF}   & \cellcolor[HTML]{EFEFEF} $\bullet$  & \cellcolor[HTML]{EFEFEF}   & \cellcolor[HTML]{EFEFEF}   & \cellcolor[HTML]{EFEFEF}   & \cellcolor[HTML]{EFEFEF} $\bullet$  & \cellcolor[HTML]{EFEFEF}   &    &    &  0.92\\
			&\cite{nirkin2020deepfake}  & 2020 &  \cellcolor[HTML]{EFEFEF}$\bullet$  & \cellcolor[HTML]{EFEFEF} $\bullet$  & \cellcolor[HTML]{DBDBDB} $\bullet$  & \cellcolor[HTML]{DBDBDB}   & \cellcolor[HTML]{DBDBDB}   & \cellcolor[HTML]{C0C0C0}   & \cellcolor[HTML]{C0C0C0}   & \cellcolor[HTML]{C0C0C0} $\bullet$  & \cellcolor[HTML]{C0C0C0} $\bullet$  &  ED-CNN  &    &  299x299  & \cellcolor[HTML]{EFEFEF}   & \cellcolor[HTML]{EFEFEF}   & \cellcolor[HTML]{EFEFEF}   & \cellcolor[HTML]{EFEFEF} $\bullet$  & \cellcolor[HTML]{EFEFEF}   & \cellcolor[HTML]{EFEFEF} $\bullet$  & \cellcolor[HTML]{EFEFEF} $\bullet$  & \cellcolor[HTML]{EFEFEF}   &    &    &  0.99\\
			&\cite{masi2020two}  & 2020 &  \cellcolor[HTML]{EFEFEF}$\bullet$  & \cellcolor[HTML]{EFEFEF} $\bullet$  & \cellcolor[HTML]{DBDBDB}   & \cellcolor[HTML]{DBDBDB} $\bullet$  & \cellcolor[HTML]{DBDBDB}   & \cellcolor[HTML]{C0C0C0}   & \cellcolor[HTML]{C0C0C0}   & \cellcolor[HTML]{C0C0C0} $\bullet$  & \cellcolor[HTML]{C0C0C0}   &  ED-LSTM  &    &  224x224  & \cellcolor[HTML]{EFEFEF}   & \cellcolor[HTML]{EFEFEF}   & \cellcolor[HTML]{EFEFEF}   & \cellcolor[HTML]{EFEFEF} $\bullet$  & \cellcolor[HTML]{EFEFEF}   & \cellcolor[HTML]{EFEFEF} $\bullet$  & \cellcolor[HTML]{EFEFEF}   & \cellcolor[HTML]{EFEFEF}   &    &    &  \\
			&\cite{wang2020cnn}  & 2020 &  \cellcolor[HTML]{EFEFEF}$\bullet$  & \cellcolor[HTML]{EFEFEF} $\bullet$  & \cellcolor[HTML]{DBDBDB} $\bullet$  & \cellcolor[HTML]{DBDBDB}   & \cellcolor[HTML]{DBDBDB}   & \cellcolor[HTML]{C0C0C0}   & \cellcolor[HTML]{C0C0C0}   & \cellcolor[HTML]{C0C0C0}   & \cellcolor[HTML]{C0C0C0} $\bullet$  &  CNN ResNet  &    &  224x224  & \cellcolor[HTML]{EFEFEF}   & \cellcolor[HTML]{EFEFEF}   & \cellcolor[HTML]{EFEFEF}   & \cellcolor[HTML]{EFEFEF} $\bullet$  & \cellcolor[HTML]{EFEFEF}   & \cellcolor[HTML]{EFEFEF}   & \cellcolor[HTML]{EFEFEF}   & \cellcolor[HTML]{EFEFEF} $\bullet$  &  Avrg.  &  Prec.=  &  0.93\\
			&\cite{guo2020fake}  & 2020 &  \cellcolor[HTML]{EFEFEF}$\bullet$  & \cellcolor[HTML]{EFEFEF} $\bullet$  & \cellcolor[HTML]{DBDBDB} $\bullet$  & \cellcolor[HTML]{DBDBDB}   & \cellcolor[HTML]{DBDBDB}   & \cellcolor[HTML]{C0C0C0}   & \cellcolor[HTML]{C0C0C0}   & \cellcolor[HTML]{C0C0C0} $\bullet$  & \cellcolor[HTML]{C0C0C0}   &  AREN-CNN  &    &  128x128  & \cellcolor[HTML]{EFEFEF}   & \cellcolor[HTML]{EFEFEF}   & \cellcolor[HTML]{EFEFEF} $\bullet$  & \cellcolor[HTML]{EFEFEF}   & \cellcolor[HTML]{EFEFEF}   & \cellcolor[HTML]{EFEFEF}   & \cellcolor[HTML]{EFEFEF}   & \cellcolor[HTML]{EFEFEF} $\bullet$  & 98.52 &    &  \\
			&\cite{mittal2020emotions}  & 2020 &  \cellcolor[HTML]{EFEFEF}$\bullet$  & \cellcolor[HTML]{EFEFEF} $\bullet$  & \cellcolor[HTML]{DBDBDB}   & \cellcolor[HTML]{DBDBDB} $\bullet$  & \cellcolor[HTML]{DBDBDB} $\bullet$  & \cellcolor[HTML]{C0C0C0} $\bullet$  & \cellcolor[HTML]{C0C0C0}   & \cellcolor[HTML]{C0C0C0} $\bullet$  & \cellcolor[HTML]{C0C0C0}   &  ED-CNN  &    &  *  & \cellcolor[HTML]{EFEFEF} $\bullet$  & \cellcolor[HTML]{EFEFEF} $\bullet$  & \cellcolor[HTML]{EFEFEF}   & \cellcolor[HTML]{EFEFEF}   & \cellcolor[HTML]{EFEFEF}   & \cellcolor[HTML]{EFEFEF}   & \cellcolor[HTML]{EFEFEF} $\bullet$  & \cellcolor[HTML]{EFEFEF}   &    &    &  0.92\\
			&\cite{agarwal2020detecting}  & 2020 &  \cellcolor[HTML]{EFEFEF}$\bullet$  & \cellcolor[HTML]{EFEFEF} $\bullet$  & \cellcolor[HTML]{DBDBDB}   & \cellcolor[HTML]{DBDBDB} $\bullet$  & \cellcolor[HTML]{DBDBDB} $\bullet$  & \cellcolor[HTML]{C0C0C0}   & \cellcolor[HTML]{C0C0C0} $\bullet$  & \cellcolor[HTML]{C0C0C0}   & \cellcolor[HTML]{C0C0C0}   &  CNN  &    &  128x128  & \cellcolor[HTML]{EFEFEF}   & \cellcolor[HTML]{EFEFEF}   & \cellcolor[HTML]{EFEFEF}   & \cellcolor[HTML]{EFEFEF}   & \cellcolor[HTML]{EFEFEF}   & \cellcolor[HTML]{EFEFEF}   & \cellcolor[HTML]{EFEFEF} $\bullet$  & \cellcolor[HTML]{EFEFEF}   & 89.6 &    &  \\
			&\cite{amerini2020exploiting}  & 2020 &  \cellcolor[HTML]{EFEFEF}$\bullet$  & \cellcolor[HTML]{EFEFEF} $\bullet$  & \cellcolor[HTML]{DBDBDB}   & \cellcolor[HTML]{DBDBDB} $\bullet$  & \cellcolor[HTML]{DBDBDB}   & \cellcolor[HTML]{C0C0C0}   & \cellcolor[HTML]{C0C0C0}   & \cellcolor[HTML]{C0C0C0} $\bullet$  & \cellcolor[HTML]{C0C0C0} $\bullet$  &  LSTM  &    &  256x256  & \cellcolor[HTML]{EFEFEF}   & \cellcolor[HTML]{EFEFEF}   & \cellcolor[HTML]{EFEFEF}   & \cellcolor[HTML]{EFEFEF} $\bullet$  & \cellcolor[HTML]{EFEFEF}   & \cellcolor[HTML]{EFEFEF}   & \cellcolor[HTML]{EFEFEF}   & \cellcolor[HTML]{EFEFEF}   & 94.29 &    &  \\
			&\cite{hsu2020deep}  & 2020 &  \cellcolor[HTML]{EFEFEF}$\bullet$  & \cellcolor[HTML]{EFEFEF} $\bullet$  & \cellcolor[HTML]{DBDBDB} $\bullet$  & \cellcolor[HTML]{DBDBDB}   & \cellcolor[HTML]{DBDBDB}   & \cellcolor[HTML]{C0C0C0}   & \cellcolor[HTML]{C0C0C0}   & \cellcolor[HTML]{C0C0C0} $\bullet$  & \cellcolor[HTML]{C0C0C0}   &  Siamese CNN  &    &  64x64  & \cellcolor[HTML]{EFEFEF}   & \cellcolor[HTML]{EFEFEF}   & \cellcolor[HTML]{EFEFEF}   & \cellcolor[HTML]{EFEFEF}   & \cellcolor[HTML]{EFEFEF}   & \cellcolor[HTML]{EFEFEF}   & \cellcolor[HTML]{EFEFEF}   & \cellcolor[HTML]{EFEFEF} $\bullet$  &  TPR=0.91  &    &  \\
			&\cite{rana2020deepfakestack}  & 2020 &  \cellcolor[HTML]{EFEFEF}$\bullet$  & \cellcolor[HTML]{EFEFEF} $\bullet$  & \cellcolor[HTML]{DBDBDB} $\bullet$  & \cellcolor[HTML]{DBDBDB}   & \cellcolor[HTML]{DBDBDB}   & \cellcolor[HTML]{C0C0C0}   & \cellcolor[HTML]{C0C0C0}   & \cellcolor[HTML]{C0C0C0} $\bullet$  & \cellcolor[HTML]{C0C0C0}   &  Ensemble  &    &  224x224  & \cellcolor[HTML]{EFEFEF}   & \cellcolor[HTML]{EFEFEF}   & \cellcolor[HTML]{EFEFEF}   & \cellcolor[HTML]{EFEFEF} $\bullet$  & \cellcolor[HTML]{EFEFEF}   & \cellcolor[HTML]{EFEFEF}   & \cellcolor[HTML]{EFEFEF}   & \cellcolor[HTML]{EFEFEF}   & 99.65 &    &  1.00\\
			&\cite{de2020deepfake}  & 2020 &  \cellcolor[HTML]{EFEFEF}$\bullet$  & \cellcolor[HTML]{EFEFEF} $\bullet$  & \cellcolor[HTML]{DBDBDB}   & \cellcolor[HTML]{DBDBDB} $\bullet$  & \cellcolor[HTML]{DBDBDB}   & \cellcolor[HTML]{C0C0C0}   & \cellcolor[HTML]{C0C0C0}   & \cellcolor[HTML]{C0C0C0} $\bullet$  & \cellcolor[HTML]{C0C0C0}   &  *  &    &  112x112  & \cellcolor[HTML]{EFEFEF}   & \cellcolor[HTML]{EFEFEF}   & \cellcolor[HTML]{EFEFEF}   & \cellcolor[HTML]{EFEFEF}   & \cellcolor[HTML]{EFEFEF}   & \cellcolor[HTML]{EFEFEF} $\bullet$  & \cellcolor[HTML]{EFEFEF}   & \cellcolor[HTML]{EFEFEF}   & 98.26 &    &  99.73\\
			&\cite{khalid2020oc}  & 2020 &  \cellcolor[HTML]{EFEFEF}$\bullet$  & \cellcolor[HTML]{EFEFEF} $\bullet$  & \cellcolor[HTML]{DBDBDB} $\bullet$  & \cellcolor[HTML]{DBDBDB}   & \cellcolor[HTML]{DBDBDB}   & \cellcolor[HTML]{C0C0C0}   & \cellcolor[HTML]{C0C0C0}   & \cellcolor[HTML]{C0C0C0} $\bullet$  & \cellcolor[HTML]{C0C0C0}   &  OC-VAE  &    &  100x100  & \cellcolor[HTML]{EFEFEF}   & \cellcolor[HTML]{EFEFEF}   & \cellcolor[HTML]{EFEFEF}   & \cellcolor[HTML]{EFEFEF} $\bullet$  & \cellcolor[HTML]{EFEFEF}   & \cellcolor[HTML]{EFEFEF}   & \cellcolor[HTML]{EFEFEF}   & \cellcolor[HTML]{EFEFEF}   &  TPR=0.89  &    &  \\
			&\cite{fernandes2020detecting}  & 2020 & \cellcolor[HTML]{EFEFEF} $\bullet$  & \cellcolor[HTML]{EFEFEF} $\bullet$  & \cellcolor[HTML]{DBDBDB} $\bullet$  & \cellcolor[HTML]{DBDBDB}   & \cellcolor[HTML]{DBDBDB}   & \cellcolor[HTML]{C0C0C0}   & \cellcolor[HTML]{C0C0C0}   & \cellcolor[HTML]{C0C0C0} $\bullet$  & \cellcolor[HTML]{C0C0C0}   &  ABC-ResNet  &    &  224x224  & \cellcolor[HTML]{EFEFEF}   & \cellcolor[HTML]{EFEFEF}   & \cellcolor[HTML]{EFEFEF}   & \cellcolor[HTML]{EFEFEF}   & \cellcolor[HTML]{EFEFEF}   & \cellcolor[HTML]{EFEFEF}   & \cellcolor[HTML]{EFEFEF}   & \cellcolor[HTML]{EFEFEF} $\bullet$  &    &  ?  &  \\

			\midrule
			& \cite{koopman2018detection} & 2018 & \cellcolor[HTML]{EFEFEF} & \cellcolor[HTML]{EFEFEF}$\bullet$ & \cellcolor[HTML]{DBDBDB} & \cellcolor[HTML]{DBDBDB}$\bullet$ & \cellcolor[HTML]{DBDBDB} & \cellcolor[HTML]{C0C0C0} & \cellcolor[HTML]{C0C0C0} & \cellcolor[HTML]{C0C0C0}$\bullet$ & \cellcolor[HTML]{C0C0C0} & PRNU &  & 1280x720 & \cellcolor[HTML]{EFEFEF} & \cellcolor[HTML]{EFEFEF} & \cellcolor[HTML]{EFEFEF} & \cellcolor[HTML]{EFEFEF} & \cellcolor[HTML]{EFEFEF} & \cellcolor[HTML]{EFEFEF} & \cellcolor[HTML]{EFEFEF} & \cellcolor[HTML]{EFEFEF}$\bullet$ & \textit{TPR=1} &\textit{\@ FPR=}  & \textit{0.03} \\
			& \cite{straub2019using} & 2019 & \cellcolor[HTML]{EFEFEF} & \cellcolor[HTML]{EFEFEF}$\bullet$ & \cellcolor[HTML]{DBDBDB} & \cellcolor[HTML]{DBDBDB}$\bullet$ & \cellcolor[HTML]{DBDBDB} & \cellcolor[HTML]{C0C0C0} & \cellcolor[HTML]{C0C0C0} & \cellcolor[HTML]{C0C0C0} & \cellcolor[HTML]{C0C0C0}$\bullet$ & Statistics & $\bullet$ & - & \cellcolor[HTML]{EFEFEF} & \cellcolor[HTML]{EFEFEF} & \cellcolor[HTML]{EFEFEF} & \cellcolor[HTML]{EFEFEF} & \cellcolor[HTML]{EFEFEF} & \cellcolor[HTML]{EFEFEF} & \cellcolor[HTML]{EFEFEF} & \cellcolor[HTML]{EFEFEF}$\bullet$ & & &  \\
			
			\multirow{-3}{*}{Statistics \& Steganalysis} & \cite{marra2019gans} & 2019 & \cellcolor[HTML]{EFEFEF}$\bullet$ & \cellcolor[HTML]{EFEFEF}$\bullet$ & \cellcolor[HTML]{DBDBDB}$\bullet$ & \cellcolor[HTML]{DBDBDB} & \cellcolor[HTML]{DBDBDB} & \cellcolor[HTML]{C0C0C0} & \cellcolor[HTML]{C0C0C0} & \cellcolor[HTML]{C0C0C0} & \cellcolor[HTML]{C0C0C0}$\bullet$ & PRNU &  & * & \cellcolor[HTML]{EFEFEF} & \cellcolor[HTML]{EFEFEF} & \cellcolor[HTML]{EFEFEF} & \cellcolor[HTML]{EFEFEF} & \cellcolor[HTML]{EFEFEF} & \cellcolor[HTML]{EFEFEF} & \cellcolor[HTML]{EFEFEF} & \cellcolor[HTML]{EFEFEF}$\bullet$ & 90.3 &  &  \\ \bottomrule\bottomrule 
			\multicolumn{26}{l}{*Only the best reported performance, averaged over the test datasets, is displayed to capture the `best-case' scenario.}
	\end{tabular}%
	}
\vspace{-2em}
\end{table*}

\egroup